\documentclass[aps,prb,twocolumn,superscriptaddress,showpacs,showkeys,amsmath,amssymb]{revtex4-1}

\usepackage{times}
\usepackage{amsfonts}
\usepackage{amssymb,amsmath}
\usepackage[dvipsnames]{xcolor}
\usepackage{graphicx}
\usepackage{bm}
\usepackage{subfigure}

\begin{document}

\title{Persistence of the flat band in a kagom\'{e} magnet with dipolar interactions}

\author{Mykola Maksymenko}
\affiliation{Institute for Condensed Matter Physics, NAS of Ukraine, 1
Svientsitskii Street, L'viv-11, Ukraine}
\affiliation{Department of Condensed
Matter Physics, Weizmann Institute of Science, Rehovot, 76100, Israel}
\author{Roderich Moessner}
\affiliation{Max-Planck-Institut f\"{u}r Physik komplexer Systeme,
        N\"{o}thnitzer Stra{\ss}e 38, 01187 Dresden, Germany}
        \author{Kirill Shtengel}
\affiliation{Max-Planck-Institut f\"{u}r Physik komplexer Systeme,
        N\"{o}thnitzer Stra{\ss}e 38, 01187 Dresden, Germany}
\affiliation{Department of Physics and Astronomy, University of California at
Riverside, Riverside,  CA 92521, USA}
\date{\today}



\begin{abstract}
The weathervane modes of the classical Heisenberg antiferromagnet on the
kagom\'{e} lattice constitute possibly the earliest and certainly the most
celebrated example of a flat band of zero-energy excitations. Such modes
arise from the underconstraint that has since become a defining criterion of
strong geometrical frustration.  We investigate the fate of this flat band
when dipolar interactions are added. These change the nearest-neighbour model
fundamentally as they remove the Heisenberg spin-rotational symmetry while
also introducing a long-range component to the  interaction. We explain how
the modes continue to remain approximately dispersionless, while being lifted
to finite energy as well as being squeezed: they change their  ellipticity
described by the ratio of the amplitudes of the canonically conjugate
variables comprising them. This phenomenon provides interesting connections
between concepts such as constraint counting and self-screening underpinning
the field of frustrated magnetism. We discuss variants of these phenomena for
different interactions, lattices and dimension.
\end{abstract}

\maketitle

\section{Introduction}
One of the hallmarks of geometrical frustration in classical spin models is a
large ground-state degeneracy associated with their ground states. For discrete
Ising spins, this leads to a non-vanishing `ground-state entropy' such as
Pauling's entropy in spin ice \cite{Ramirez1999}, for a review
see~\cite{Castelnovo2012}. For continuous spins, the ground states form a
manifold of extensive dimensionality~\cite{Moessner1998a,Moessner1998b}.
Continuous rearrangements of spins effecting moves between these degenerate
ground-state configurations form zero-energy modes of the system.

If such zero-energy modes are local, i.e.\ involve only a finite number of
spins, they are often referred to as weathervane
modes\cite{Chalker1992,Ritchey1993,Shender1993}. Their locality implies a
straightforward possibility of a non-zero  density of them, and  hence an
extensive number resulting in the existence of a flat zero-energy band in
momentum space. A typical scenario for the appearance of such modes in
classical spin models is provided by the locally underconstrained nature of the
spin interactions, whereby a mismatch between the number of degrees of freedom
and the number of ground state constraints imposed by the Hamiltonian is
extensive~\cite{Moessner1998a,Moessner1998b}. As such, the existence of these
local modes is not protected by any symmetries and can be easily destroyed by
perturbations -- e.g. next-nearest neighbour interactions -- which typically
increase the number of constraints, rendering the  formerly flat zero-energy
bands dispersive~\cite{Harris1992}. When such additional interactions also
break the symmetries of the Hamiltonian, even gaplessness of the spectrum is no
longer guaranteed  and  interesting phenomenon can arise: the formerly flat
zero-energy bands can be lifted up to finite energy but remain
flat~\cite{Chernyshev2015}. One example of such behaviour has been recently
observed in the kagom\'{e} Heisenberg antiferromagnet (KHAFM) with an
additional out-of-plane Dzyaloshinskii-Moriya (DM) interaction (albeit in this
case another magnon band remained gapless due to the remaining U(1) symmetry of
the Hamiltonian). The persisting flatness of such a band implies the local
nature of the spin excitations; in other words such excitations preserve their
weathervane character.

Many recent discussions have been centred around flat bands in fermionic
systems due to the fact that, if partially occupied, they provide a fertile
ground for a variety of unconventional orders (for reviews, see e.g.
Ref.~\onlinecite{Parameswaran2013a,Bergholtz2013,Derzhko2015}).  More exotic
flavours and settings--e.g.\ lattices for which all bands are
flat~\cite{Vidal1998}, lattices with `higher-spin' Weyl
fermions~\cite{Dora2011}, or in the currently very fashionable Floquet
setting~\cite{Du2017} -- have also been considered over the years. Even when
not hosting any  exotic topological orders, such bands can nevertheless be
responsible for unusual thermodynamic, dynamic and transport properties. These
properties need not be restricted to fermionic flat bands; their bosonic
counterparts can be responsible for a variety of interesting phenomena such as
magnetisation plateaux in frustrated magnets at high magnetic
fields~\cite{Schulenburg2002,Zhitomirsky2005,Schmidt2006,Derzhko2007a}.

In this work we report a simple example of a system hosting a flat magnon band
in the absence of any external field: a large-spin KHAFM with additional
dipolar spin--spin interaction. The dipolar interaction breaks the SU(2)
symmetry of the Heisenberg Hamiltonian down to $\mathbb{Z}_2\times
\mathbb{Z}_2$, the largest subgroup consistent with the point group symmetry of
the kagom\'{e} lattice, and hence (unsurprisingly) completely gaps the magnon
spectrum. What \emph{is} surprising is that the lowest excitation band stays
perfectly flat in the case of truncated nearest-neighbour dipolar interactions
and remains flat to a very good approximation for the complete long-range case.
The energy of the flat band is proportional to $\sqrt{\gamma}$, where $\gamma$
is the strength of the dipolar term. While in the case of constraint counting
for the nearest-neighbour KHAFM, the flatness simply follows from the fact that
the modes are pinned to zero energy, its origin here is considerably less
transparent.

Our central result is that the survival of flatness rests on two ingredients.
The first is that the canonical structure of the flat modes comprises a pair of
conjugate variables, \emph{both of which} have a flat momentum dependence, so
that they can combine into a flat band at finite frequency. This double
flatness is a very peculiar feature of the kagom\'{e} magnet which, to the best
of our knowledge, has not yet been explicitly identified. As the magnitude of
the dipolar term is increased, the ellipticity of the mode changes from 0 for
the Heisenberg weathervane mode to a finite value, reminiscent to the physics
of squeezing in quantum optics, but with a concomitant change in frequency for
the modes here.

For the case of truncated dipolar interactions with their range limited to the
nearest neighbours, we present an exact analytical calculation illuminating the
microscopic nature of the modes comprising the flat band. The survival of the
flatness for the full long-range dipolar interaction requires the second
ingredient: both conjugate variables encode a mode with vanishing net spin
moment. This renders the long-range (inverse cube) tails of the dipolar
interaction ineffective for this particular type of spin excitations, the long
corrections are dominated by shorter range, higher terms in the multipole
expansion. This is analogous to the phenomenon of self-screening first noticed
for Ising spins in the context of dipolar spin ice\cite{denHertogGingras:2000}
and later explained in terms of a projective equivalence\cite{isakov:2005},
with a simple picture provided by a dumbbell model for the Ising spins
\cite{Castelnovo2008}; our model provides an example of such a behaviour for
continuous spins.

Furthermore, we show that the presence of a flat spin wave mode is a rather
generic feature of a large spin kagom\'{e} antiferromagnet. This effect is
found to survive the addition of other interaction terms such as the
Dzyaloshinskii--Moriya or anisotropic terms (consistent with the symmetries of
the kagom\'{e} lattice) within an extended range of parameters.

We note that finite frequency flat bands have been experimentally observed in
other frustrated spin systems with dipolar interactions, such as gadolinium
gallium garnet~\cite{dAmbrumenil2015}; the physics described here need not be
limited to the kagom\'{e} lattice.

\section{2D kagom\'{e} Heisenberg antiferromagnet}
\label{sec:KHAFM}

In this section we briefly review some known results for the classical KHAFM.
The Heisenberg Hamiltonian can be written as
\begin{equation}\label{eq:Heisenberg_ham}
H_\text{KHAFM}=J\sum_{\langle i,j\rangle} \mathbf{S}_i \cdot \mathbf{S}_j
= \frac{J}{2}\sum_\vartriangle \mathbf{S_\vartriangle}^2
+ \text{const}
\end{equation}
where the second sum is performed over all triangles of the kagom\'{e} lattice and
$\mathbf{S_\vartriangle}$ is the combined spin of the three sites forming a
given triangle. The energy is minimised by a state with all
$\mathbf{S_\vartriangle}=0$, implying that any $120^\circ$ arrangement of spins
around every triangle corresponds to a classical ground state of this
Hamiltonian.

\begin{figure}[htb]
\begin{center}
    \subfigure[]{\includegraphics[width=0.33\textwidth]{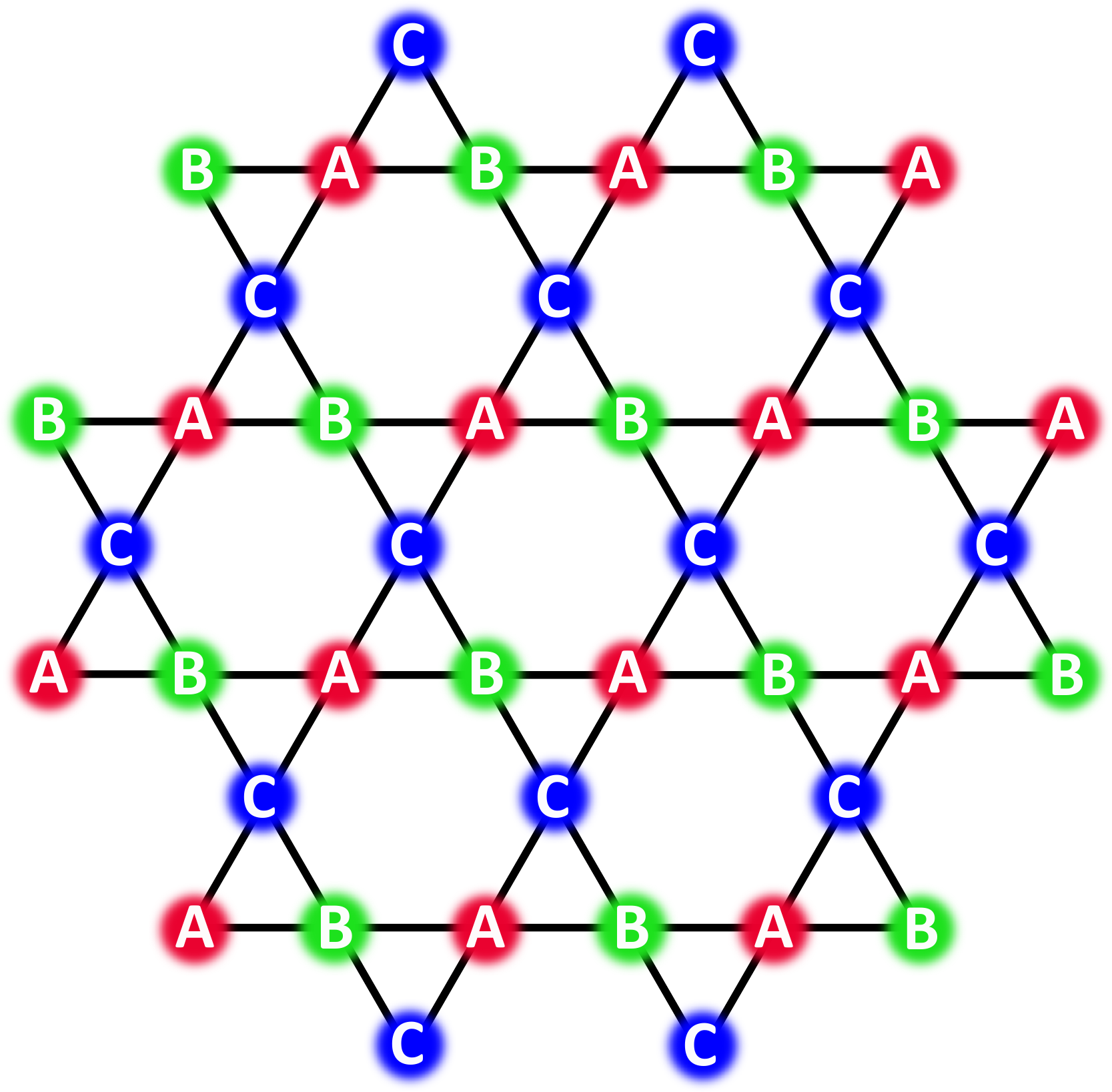}
    \label{fig:q=0}}
    \subfigure[]{\includegraphics[width=0.33\textwidth]{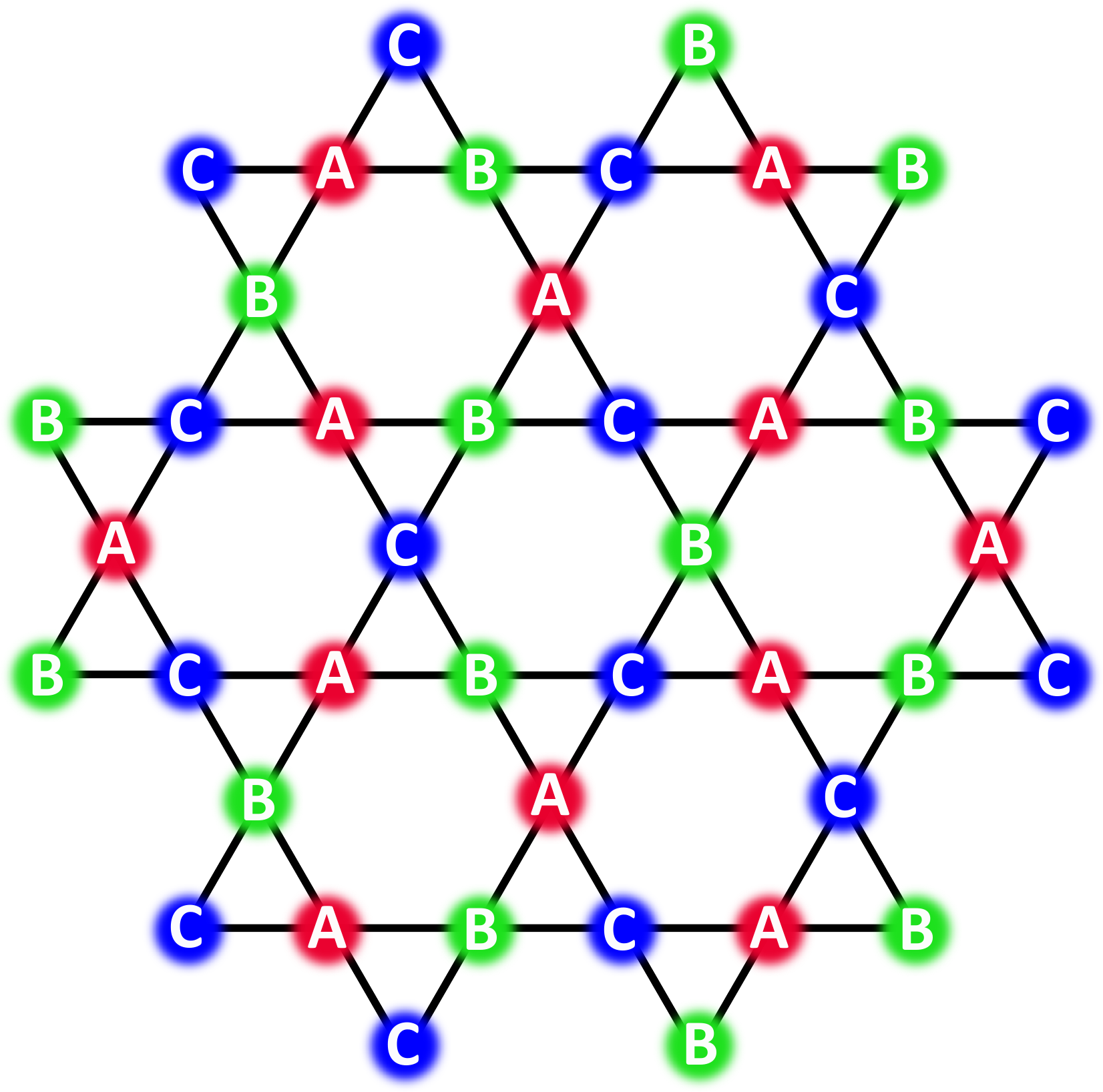}
    \label{fig:root_3}}
    \caption{(Colour online) Ground states of the KHAFM: (a) the $q=0$ state;
    (b) the $\sqrt{3}\times\!\sqrt{3}$ state. A, B and C here label three
    directions of spins in a $120^\circ$ configuration:
    $\mathbf{S}_\text{A} + \mathbf{S}_\text{B} + \mathbf{S}_\text{C}=0$.}
\label{fig:KHAFM_gs}
\end{center}
\end{figure}

The ground state manifold has an extensive dimensionality. One of the key
results of the early spin-wave analyses of this model is that a co-planar
subset of this manifold is selected by the order-by-disorder
mechanism~\cite{Chalker1992,Harris1992,Huse1992,Ritchey1993}.
The two most prominent types of these states, the $q=0$ and
$\sqrt{3}\times\!\sqrt{3}$ states, are shown here in Figure~\ref{fig:KHAFM_gs}.
In linear spin-wave theory, the spectra of all coplanar states are
identical\cite{Chalker1992}, a feature not present in the full nonlinear
problem~\cite{Chubukov1992,von_Delft1992,Chern2013}.

The $\sqrt{3}\times\!\sqrt{3}$ state, shown in
Figure~\ref{fig:root_3}, serves as an archetypal example of a state supporting
an exact local zero energy mode -- a.k.a.\  weathervane mode -- which
corresponds to rotations of e.g. A- and B-spins located around a single hexagon
of the lattice around the axis given by the direction of the C-spins.
Nevertheless \emph{all} co-planar ground states posses such soft  modes
in the harmonic approximation, and hence are characterised by a zero-energy
magnon band.

\section{Adding dipolar interactions}
\label{sec:dipolar_interactions}

We now extend the KHAFM model (\ref{eq:Heisenberg_ham}) by adding magnetic
dipolar interations between spins:
\begin{equation}
H_\text{KH+D}=J\sum_{\langle i,j\rangle} \mathbf{S}_i \cdot \mathbf{S}_j
+\frac{1}{2}\sum_{ i,j} \gamma_{ij}^{\alpha\beta}S_{i}^{\alpha}S_{j}^{\beta}
\label{eq:dipolar_hamiltonian}
\end{equation}
where the second sum is not restricted to nearest neighbours and
\begin{equation}
\gamma_{ij}^{\alpha\beta}= \gamma a^3
\left(\frac{\delta_{\alpha\beta}}{\lvert{\mathbf{r}_{ij}}\rvert^{3}}
-3\frac{{r}_{ij}^{\alpha}{r}_{ij}^{\beta}}{\lvert{\mathbf{r}_{ij}}\rvert^{5}}\right).
\label{eq:int_matrix}
\end{equation}
Here $a$ is the lattice constant, $\mathbf{r}_{ij}\equiv \mathbf{r}_{j}
-\mathbf{r}_{i}$ and $\gamma = g^2 \mu_\text{B}^2/a^3$ (in Gaussian units).

\begin{figure}[htb]
\begin{center}
\includegraphics[width=0.3\textwidth]{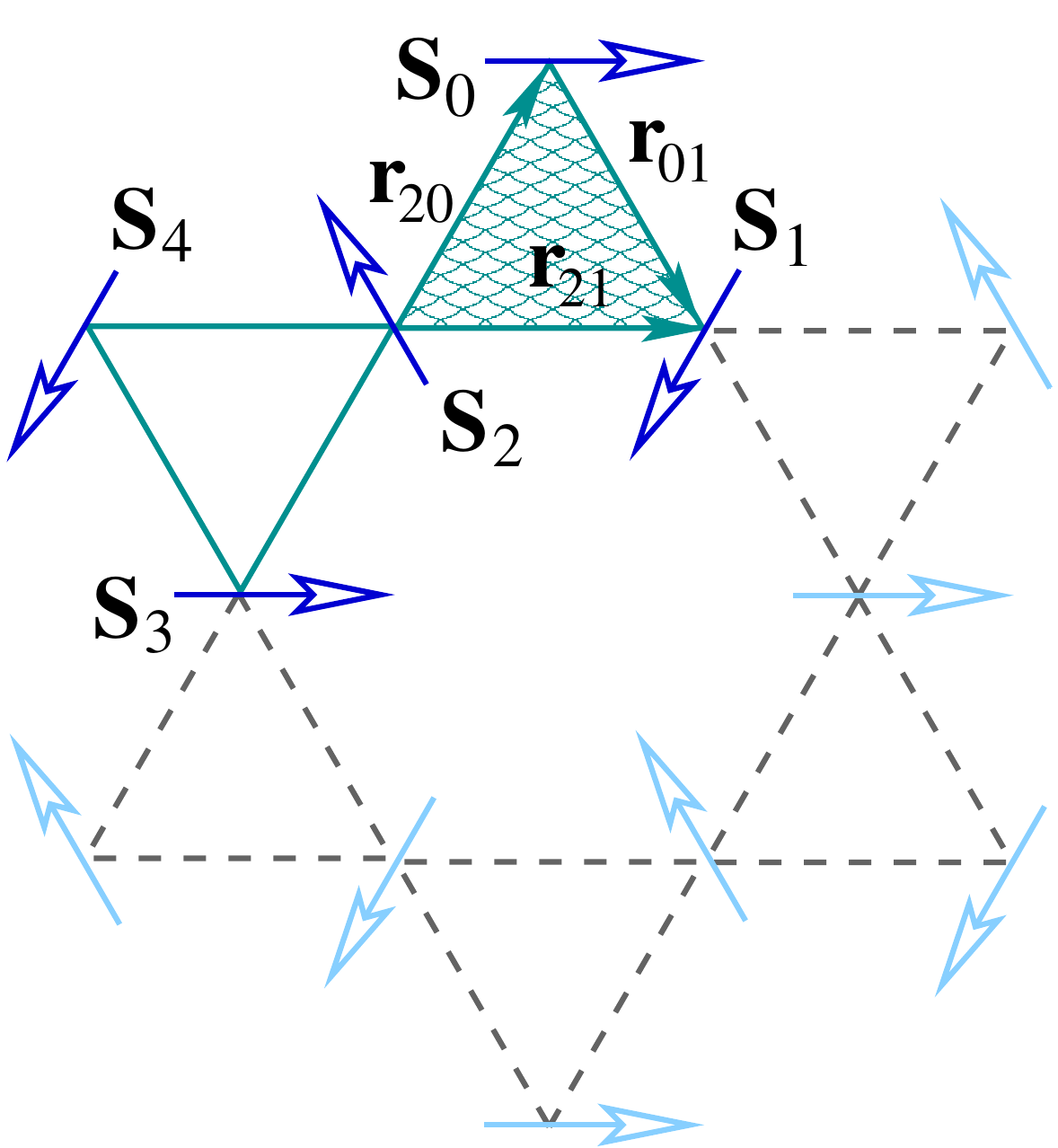}
\caption{(Colour online) One of the two classical ground states of the KHAFM
with the additional dipolar interactions between spins.}
\label{fig:dipolar_gs}
\end{center}
\end{figure}

Remarkably, in the presence of a relatively small Heisenberg
term, this dipolar term does not compete with the nearest-neighbour exchange:
the ground state is still one the of $120^\circ$
states. More specifically, the ground state, shown here in
Figure~\ref{fig:dipolar_gs}, is one of the $q=0$ states (see
Figure~\ref{fig:q=0}). The spins at each site align in the direction parallel
to that of the opposite side of a triangle this spin belongs to, with the three
spins around each triangle adding to 0. This ground state is doubly degenerate,
with respect to a global reversal of all spins.

Note that the $120^{^\circ}$ state is \emph{not} a ground state of the long-range dipolar
Hamiltonian alone~\cite{Maksymenko2015b}. However, it is easy to verify its local
\emph{extremality}. Let us
first consider the case of dipolar interactions whose range is restricted only
to the nearest neighbours. The Hamiltonian is
\begin{multline}
H_\text{KH+NND}=\sum_{\langle i,j\rangle} \big\{(J+\gamma) \mathbf{S}_i \cdot \mathbf{S}_j
\\
-3 \gamma \left(\mathbf{S}_i\cdot \hat{\mathbf{r}}_{ij}\right)
\left(\mathbf{S}_j\cdot \hat{\mathbf{r}}_{ij}\right)
\big\},
\label{eq:nn_dipolar_hamiltonian}
\end{multline}
where $\hat{\mathbf{r}}_{ij}$ is the unit vector directed from site $i$ to site
$j$. In order to analyse the ground state of this Hamiltonian, let us focus on
just two of its terms, specifically those involving interactions between spin
$\mathbf{S}_0$ and two of its neighbours, $\mathbf{S}_1$ and $\mathbf{S}_2$,
shown in Figure~\ref{fig:dipolar_gs}. These terms can be rewritten as
\begin{multline}
H_{01} + H_{02} \\
= \big\{(J+\gamma) \mathbf{S}_0 \cdot \mathbf{S}_1
-3 \gamma \left(\mathbf{S}_0\cdot \hat{\mathbf{r}}_{01}\right)
\left(\mathbf{S}_1\cdot \hat{\mathbf{r}}_{01}\right)
\big\}\\
+ \big\{(J+\gamma) \mathbf{S}_0 \cdot \mathbf{S}_2
-3 \gamma \left(\mathbf{S}_0\cdot \hat{\mathbf{r}}_{20}\right)
\left(\mathbf{S}_2\cdot \hat{\mathbf{r}}_{20}\right)
\big\}\\
= - \mathbf{\tilde{h}}_0\cdot\mathbf{S}_0,
\label{eq:012_dipolar_hamiltonian}
\end{multline}
where
\begin{multline}\label{eq:effective_field}
 \mathbf{\tilde{h}}_0 = - (J+\gamma)\left( \mathbf{S}_1  + \mathbf{S}_2\right)
 \\
 + 3\gamma \left[\hat{\mathbf{r}}_{01}\left(\mathbf{S}_1\cdot \hat{\mathbf{r}}_{01}\right)
 + \hat{\mathbf{r}}_{20}
 \left(\mathbf{S}_2\cdot \hat{\mathbf{r}}_{20}\right)\right]\\
 = \hat{\mathbf{r}}_{21}\left(J+5\gamma/2\right)S.
\end{multline}
In deriving the last line we used the fact that in the state shown in
Figure~\ref{fig:dipolar_gs}, $\mathbf{S}_1  + \mathbf{S}_2 = - \mathbf{S}_0 = -
S\hat{\mathbf{r}}_{21}$ and $\mathbf{S}_1\cdot \hat{\mathbf{r}}_{01} =
\mathbf{S}_2\cdot \hat{\mathbf{r}}_{20} = S/2$ as well as
$\hat{\mathbf{r}}_{20} + \hat{\mathbf{r}}_{01} =\hat{\mathbf{r}}_{21}$. In
other words, spin $\mathbf{S}_0$ is aligned with the effective magnetic field
due to its neighbours (the effective field is actually $\mathbf{h}_0
=2\mathbf{\tilde{h}}_0$ since another $\mathbf{\tilde{h}}_0$ results from
$\mathbf{S}_0$'s two other neighbours).

Next, consider the following two observations. Firstly, accounting for
longer-range dipolar interactions does not tilt $\mathbf{h}_0$ away from the
horizontal, albeit can potentially reverse its direction (thus making this
equilibrium state unstable). In order to see this, consider a pair of spins
situated on the same horizontal line at sites which are symmetric with respect
to the vertical line passing through the location of $\mathbf{S}_0$ (see
Figure~\ref{fig:dipolar_gs}). Using the fact that this state is a $q=0$ state
shown in Figure~\ref{fig:q=0}, we know that either both of the spins are of
type C (i.e., the same as $\mathbf{S}_0$) or one of them is of type A while the
other one is of type B. Denoting these sites as $k$ and $l$, we note that
either $\mathbf{S}_k+ \mathbf{S}_l = 2\mathbf{S}_0$ (if both $\mathbf{S}_k$ and
$\mathbf{S}_l$ are of type C) or else $\mathbf{S}_k+ \mathbf{S}_l =
-\mathbf{S}_0$. Meantime $\mathbf{S}_k\cdot \hat{\mathbf{r}}_{0k} =
\mathbf{S}_l\cdot \hat{\mathbf{r}}_{l0}$ while $\hat{\mathbf{r}}_{l0} +
\hat{\mathbf{r}}_{0k} \propto \hat{\mathbf{r}}_{lk}$. In other words, the
contribution of the pair of spins at sites $k$ and $l$ into the effective field
acting on $\mathbf{S}_0$ is still collinear with the direction of
$\mathbf{S}_0$ albeit its sign depends on the location of the pair. As a
result, the net effective field due to all spins interacting with
$\mathbf{S}_0$ via untruncated dipolar coupling need not be positive. In fact,
that is exactly what happens, and the state shown in
Figure~\ref{fig:dipolar_gs} is not a ground state of long-range dipolar
Hamiltonian alone~\cite{Maksymenko2015b}. However, a small nearest-neighbour
Heisenberg term $J\gtrsim 0.1 \gamma$ (which is already accounted for in
Eq.~(\ref{eq:effective_field})) will align $\mathbf{h_0}$ with $\mathbf{S}_0$
and stabilise this ground state~\cite{Maksymenko2015b}.

Appendix~\ref{sec:dipolar_alt} presents a different way of writing the
Hamiltonian for the nearest neighbour Heisenberg and dipolar interactions which
is useful in analysing the stability of the aforementioned ground state in the
presence of some other interactions, such as the Dzyaloshinskii--Moriya
interaction.

The stability analysis presented here is also a useful departing point for
understanding the nature of the flat spin-wave band in this system. In
particular, it allows for an identification of a local mode in the case of
nearest-neighbour dipolar interactions. Specifically, we note that according to
Eq.~(\ref{eq:effective_field}), the direction of the ``restoring field''
$\mathbf{h}_0$ acting on $\mathbf{S}_0$ due to the two neighbouring spins,
$\mathbf{S}_1$ and $\mathbf{S}_2$, remains unchanged if these two spins tilt
away from their equilibrium positions by the same angle in  opposite
directions. This is because $\mathbf{S}_1 + \mathbf{S}_2$ remains parallel to
$\hat{\mathbf{r}}_{21}$ while $\mathbf{S}_2\cdot
\hat{\mathbf{r}}_{20}=\mathbf{S}_1\cdot \hat{\mathbf{r}}_{01}$ . Therefore,
spin $\mathbf{S}_0$ experiences no torque and maintains its orientation.
Moreover, if all six spins around a given hexagon (e.g. $\mathbf{S}_1$,
$\mathbf{S}_2$, $\mathbf{S}_3 \ldots$ in Fig.~\ref{fig:dipolar_gs}) rotate by
the same amounts in  alternating directions, none of the surrounding spins
(i.e. $\mathbf{S}_0$, $\mathbf{S}_4 \ldots$ in Fig.~\ref{fig:dipolar_gs}) would
experience any torque as a result of such a vibrational mode -- i.e., the mode
will remain local.

In the next section we provide a semiclassical treatment of such a mode to show
that it is indeed an eigenmode of the
Hamiltonian~(\ref{eq:nn_dipolar_hamiltonian}) and evaluate its frequency.

\section{The weathervane mode}
\label{sec:local_mode}

\begin{figure}[htb]
\centering
\includegraphics[trim=12cm 18cm 15.4cm 10cm, clip=true, width=\columnwidth]{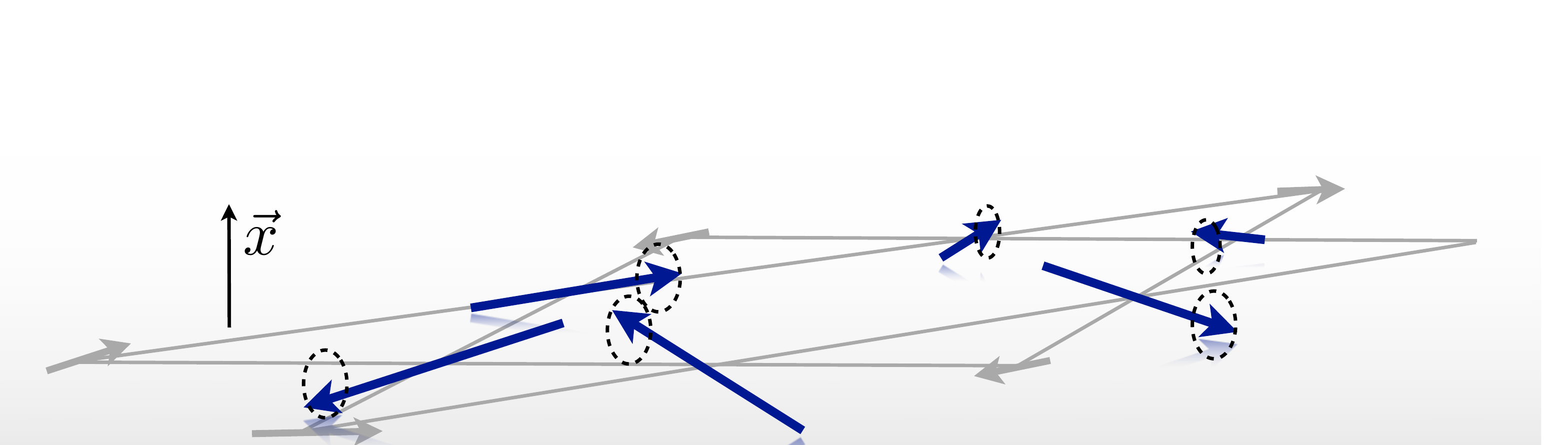}
\includegraphics[trim=0cm 0cm 0cm 0cm, clip=true, width=\columnwidth]{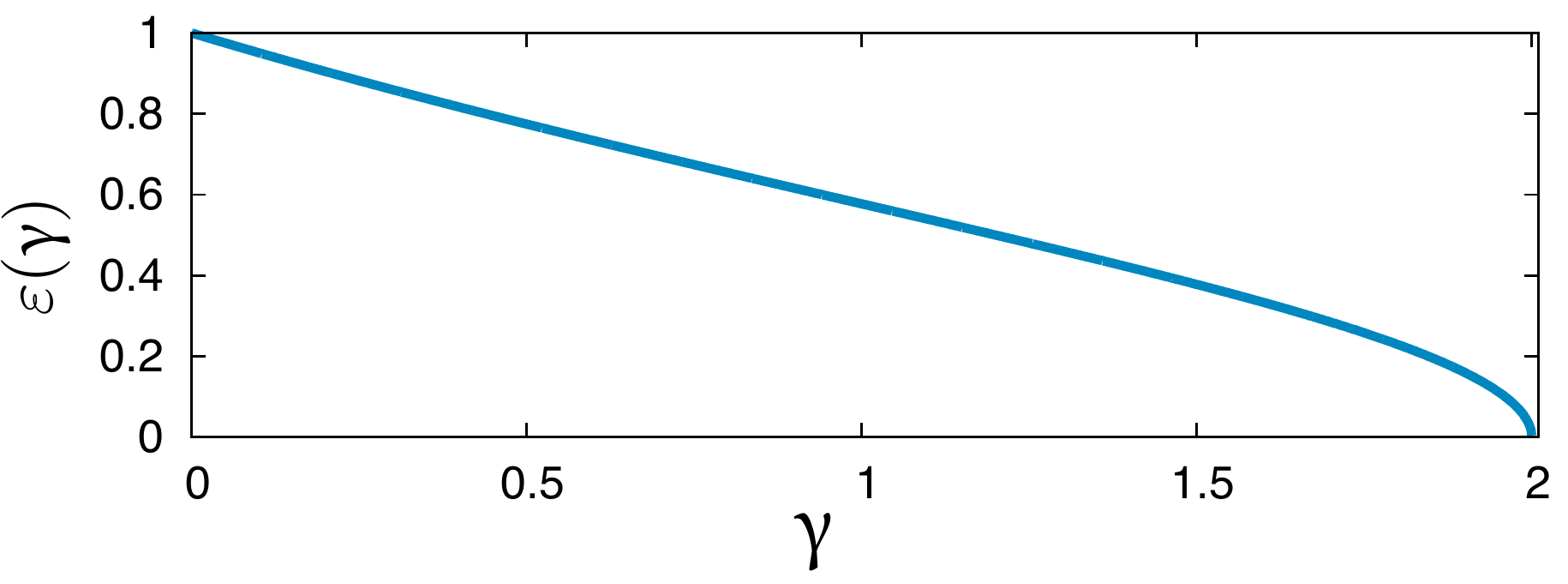}
\caption
{(Colour online) In the presence of nearest-neighbour dipolar interactions,
the weathervane mode of KHAFM remains perfectly local,
but the spin precession becomes elliptic with eccenticity $\varepsilon(D)$
given by Eq.~(\ref{eq:eccentricity}) and plotted here. }
\label{fig:lattice}
\end{figure}

Having identified the nature of the local mode, we  now proceed with the
semiclassical equations of motions for the affected spins. To that end, we
shall focus on spin $\mathbf{S}_2$. Provided that all terms in the Hamiltonian
containing $\mathbf{S}_2$ can be combined together so that $H_2 = -
\mathbf{h}_2\cdot \mathbf{S}_2$, the EOM for this spin is given by
\begin{equation}
\label{eq:Landau_Lifshitz}
\hbar \frac{d \mathbf{S}_2}{dt}=\mathbf{S}_2\times\mathbf{h}_2,
\end{equation}
which is simply the Landau--Lifshitz equation. Here $\mathbf{h}_2$ is the
effective field due to the interaction of $\mathbf{S}_2$ with four neighbouring
spins; we can readily use Eq.~(\ref{eq:effective_field}) to write
\begin{multline}\label{eq:effective_field_b}
 \mathbf{h}_2
 = - (J+\gamma)\left(\mathbf{S}_0+ \mathbf{S}_1  + \mathbf{S}_3+ \mathbf{S}_4\right)\\
 + 3\gamma \left[\hat{\mathbf{r}}_{20} \left(\mathbf{S}_0\cdot \hat{\mathbf{r}}_{20}\right)
 + \hat{\mathbf{r}}_{21}\left(\mathbf{S}_1\cdot \hat{\mathbf{r}}_{21}\right)\right.\\
 \left. + \hat{\mathbf{r}}_{23}\left(\mathbf{S}_3\cdot \hat{\mathbf{r}}_{23}\right)
 + \hat{\mathbf{r}}_{24}\left(\mathbf{S}_4\cdot \hat{\mathbf{r}}_{24}\right)
 \right].
\end{multline}
Using the fact that spins $\mathbf{S}_0$ and $\mathbf{S}_4$ keep their ground
state orientation, we can simplify this equation to read
\begin{multline}\label{eq:effective_field_c}
\mathbf{h}_2
 = - \left(J+{5}\gamma/2\right)S \hat{\mathbf{r}}_{01}
 -(J+\gamma)\left(\mathbf{S}_1+\mathbf{S}_3\right)\\
 + 3\gamma \left[
  \hat{\mathbf{r}}_{21}\left(\mathbf{S}_1\cdot \hat{\mathbf{r}}_{21}\right)
 + \hat{\mathbf{r}}_{23}\left(\mathbf{S}_3\cdot \hat{\mathbf{r}}_{23}\right)
 \right].
\end{multline}
Let us choose local right-handed coordinate systems so that the $z$-axis at
each site points along the equilibrium direction of its spin while all $x$-axes
point out of the plane (towards the reader). Recall that in the putative local
weathervane mode, spins $\mathbf{S}_1$ and $\mathbf{S}_3$ tilt away from their
equilibrium positions by the same amount. Thus the respective components of
$\mathbf{S}_1$ and $\mathbf{S}_3$ are equal to one another,
$S_1^\alpha=S_3^\alpha$  (which, of course \emph{does not} imply that
$\mathbf{S}_1= \mathbf{S}_3$ since the local axes are different at these
sites!). We can then write the effective field at the location of
$\mathbf{S}_2$ as
\begin{equation}
\mathbf{h}_2 = \left(\!-2 \left( J\!+\!\gamma\right)\!S^{x}_{1},
\left(\!J\!-\!\frac{7\gamma}{2}\!\right)\!S^{y}_{1}, \left(\!J+\frac{5}{2} \gamma\!\right)\left(S^{z}_{1}+S\right)\!\right)\nonumber
\end{equation}
and the resulting linearised EOMs then become
\begin{subequations}\label{eq:EOMs}
\begin{equation}\label{eq:EOMx}
\hbar \frac{d S^{x}_{2}}{d t}
\approx S\left(\left(5\gamma+2J\right)S^{y}_{2}-\left(J-\frac{7}{2}\gamma\right)S^{y}_{1}\right)
\end{equation}
\begin{equation}\label{eq:EOMy}
\hbar \frac{d S^{y}_{2}}{d t} 
\approx S \left(-2\left( J+\gamma\right)
S^{x}_{1}-\left(2J+5\gamma\right)S^{x}_{2}\right),
\end{equation}
\begin{equation}\label{eq:EOMz}
\hbar\frac{dS^{z}_{2}}{dt} = S^{x}_{2}h^{y}-S^{y}_{2}
h^{x}=\mathcal{O}({S^{x}}^2,{S^{y}}^{2})\approx 0,
\end{equation}
\end{subequations}
where we used the fact that $S_i^z = S +
\mathcal{O}({S^{x}}^2,{S^{y}}^{2})\approx S$. We can further simplify these
equations if we recall that the neighbouring spins participating in the
putative weathervane mode tilt in opposite directions, and therefore
$S^{x}_{1}=-S^{x}_{2}$, $S^{y}_{1}=-S^{y}_{2}$. The linearised EOMs finally
read
\begin{equation}\label{eq:EOM_lin}
  \frac{\hbar}{S} \frac{d S^{x}_{2}}{d t} =  3\left( J+\frac{\gamma}{2}\right) S^{y}_{2},
  \qquad
  \frac{\hbar}{S} \frac{d S^{y}_{2}}{d t} =  - 3\gamma S^{x}_{2}.
\end{equation}
From here, the the energy of this mode is
\begin{equation}\label{eq:freq}
  {\hbar}\omega =  3 S\sqrt{\gamma\left( J+\frac{\gamma}{2}\right)}.
\end{equation}
For $\gamma\ll J$, $\omega\propto \sqrt{\gamma}$. The tip of each of the six
spins around a hexagon traces an ellipse around its equilibrium position, with
the eccentricity parameter given by
\begin{equation}
\varepsilon = \sqrt{\frac{2J-\gamma}{2J+\gamma}}
\label{eq:eccentricity}
\end{equation}
and shown in Fig.~\ref{fig:lattice}.

\section{Spin-wave approach}

\begin{figure}[hbt]
  \begin{center}
\subfigure[]{\label{fig:spin_wave_spectrum_lr}
\includegraphics[trim=1.cm 2.4cm 1.cm 2.5cm, clip=true,width=0.44\textwidth]{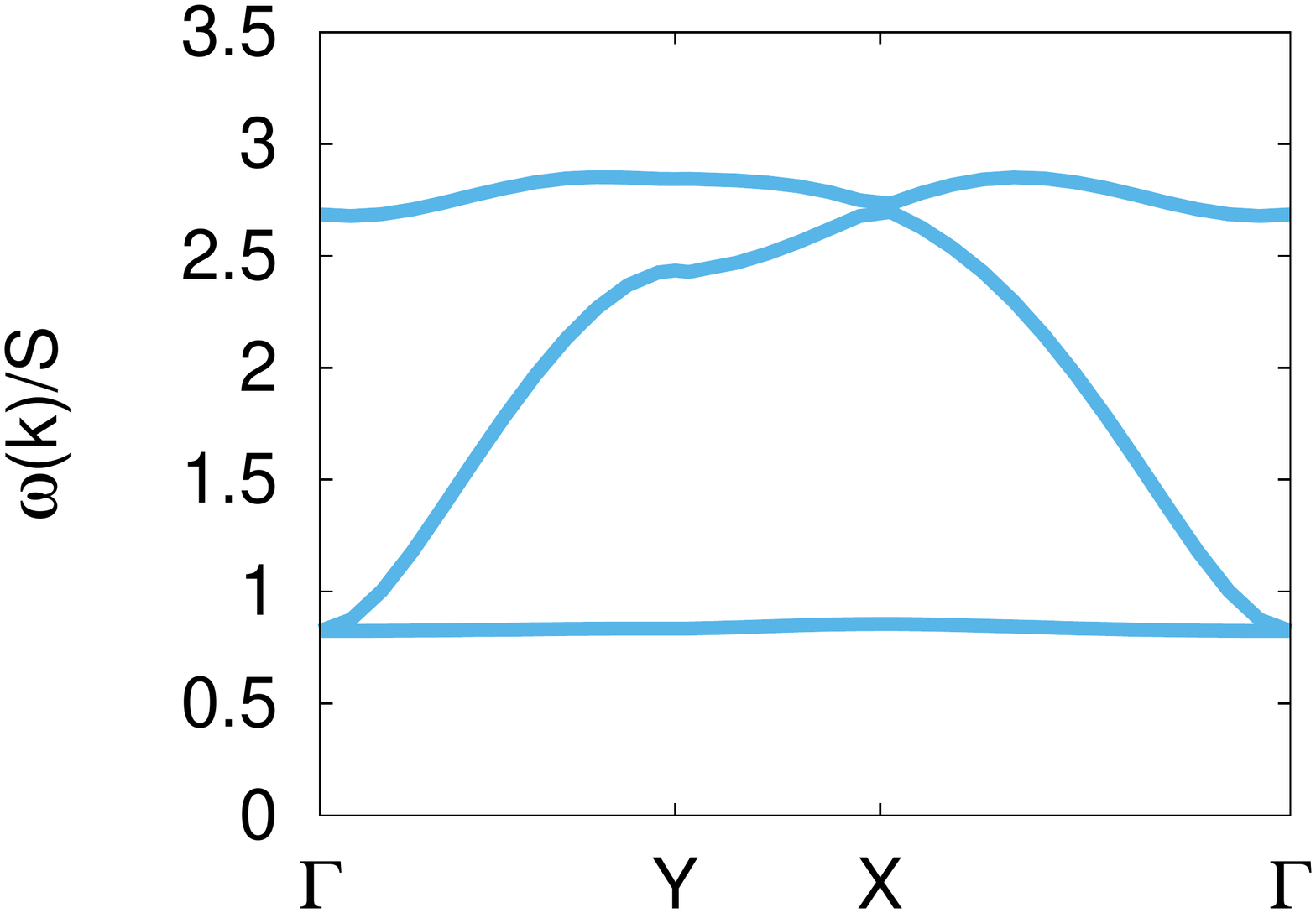}}
\subfigure[]{\label{fig:spin_gap_scaling}
\includegraphics[trim=1.8cm 2.4cm 1.cm 2.5cm, clip=true,width=0.44\textwidth]{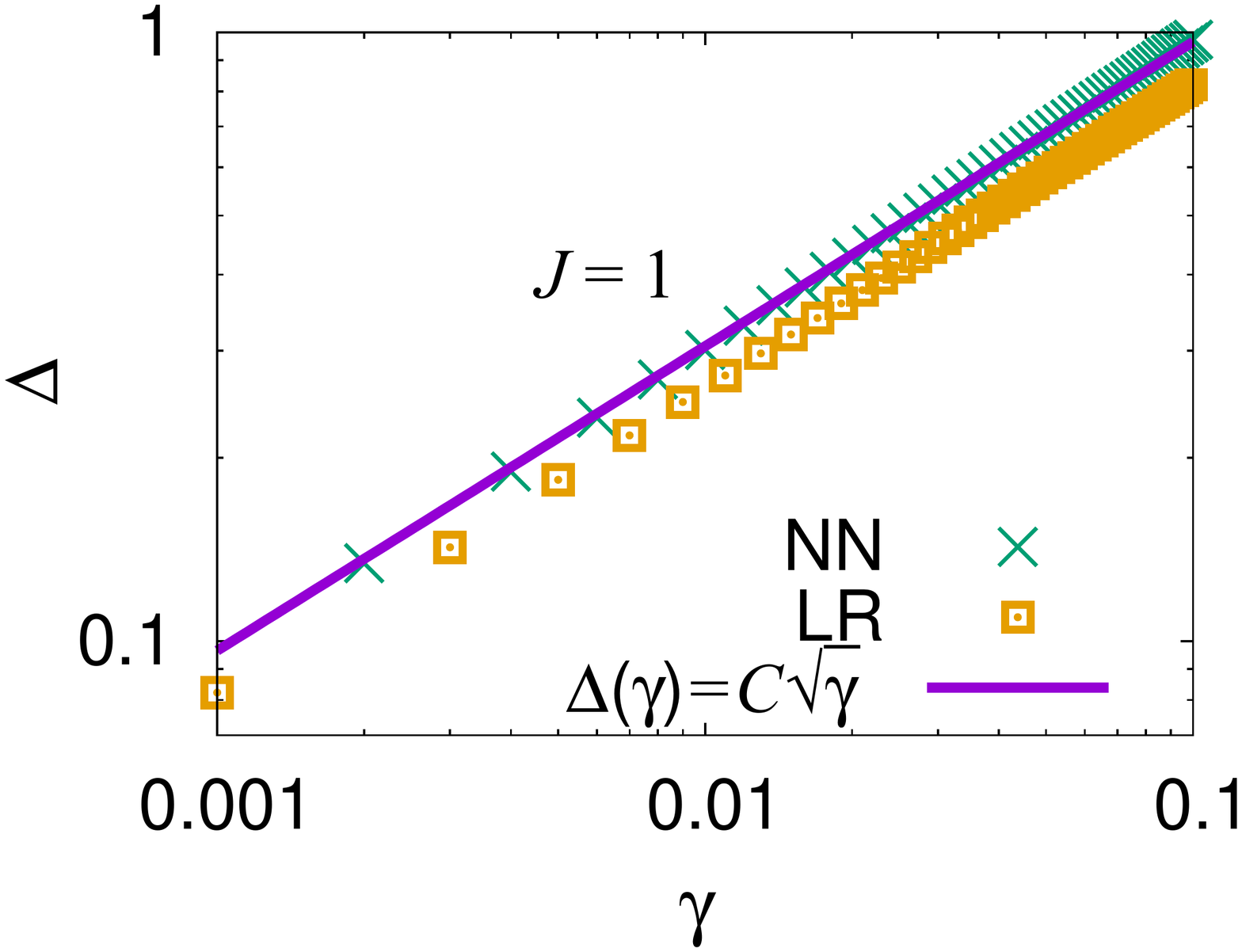}}
  \end{center}
\caption{(Colour online) (a) Spin-wave spectrum of kagome antiferromagnet with nearest neighbour
Heiseberg ($J=1$) and $\gamma=0.1 J$ long-range dipolar interactions.
The dipolar interactions gap the spectrum, but the lowest band remains nearly dispersionless.
(b) The gap scales as $\sqrt{\gamma}$ both for long-range (LR) and truncated nearest-neighbour (NN) dipolar interactions. }
\label{fig:spin_wave_spectrum}
\end{figure}

The findings of the previous section can be confirmed in the spin-wave theory
and extended to long-range dipolar interactions. The quantum fluctuations
around the classical ground state are naturally obtained after the linearised
Holstein--Primakoff transformation
\begin{eqnarray}
{S_{i}^{x}}(\mathbf{k}) & = & \sqrt{\frac{S}{2}}\left[c_{i}^{\dagger}(\mathbf{k})+c_{i}(-\mathbf{k})\right]\\
{S_{i}^{y}}(\mathbf{k}) & = & i\sqrt{\frac{S}{2}}\left[c_{i}^{\dagger}(\mathbf{k})-c_{i}(-\mathbf{k})\right]\nonumber\\
{S_{i}^{z}}(\mathbf{k}) & = & \sqrt{N}S\delta_{\mathbf{k},0}\,e^{-i\mathbf{k}\cdot\mathbf{r}_{i}}-\frac{1}{\sqrt{N}}\sum_{\mathbf{k}^{\prime}}c_{i}^{\dagger}(\mathbf{k}^{\prime})c_{i}(\mathbf{k}^{\prime}-\mathbf{k}),\nonumber
\end{eqnarray}
with boson operators  $\left[c_{i}(\mathbf{k}),\;
c_{j}^{\dagger}(\mathbf{k}^{\prime})\right]=\delta_{i,j}\delta_{\mathbf{k},\mathbf{k^{\prime}}}$.
As before, the components of the spin vector are obtained in each spin's local
coordinate frame.
Truncating the Hamiltonian beyond the quadratic terms and diagonalizing the
resulting quadratic form by the means of a standard Bogolyubov transformation,
we arrive at the  Hamiltonian
\begin{eqnarray}
H&=&H^{(0)}+\sum_{\mathbf{k}}\sum_{i}\omega_{i}(\mathbf{k})\\
&&+\sum_{\mathbf{k}}\sum_{i}\omega_{i}(\mathbf{k})\left[a_{i}^{\dagger}(\mathbf{k})a_{i}(\mathbf{k})+a_{i}^{\dagger}(-\mathbf{k})a_{i}(-\mathbf{k})\right],\nonumber
\end{eqnarray}
where $a_{i}(\mathbf{k})$ and $a^{\dagger}_{i}(\mathbf{k})$ are
Bogolyubov-transformed boson operators and the real eigenvalues
$\omega_{i}(\mathbf{k})$ indicate stable
ground-state spin configuration.

\begin{figure}
\includegraphics[trim=2.cm 2.cm 1.8cm 2.cm, clip=true,width=0.45\textwidth]{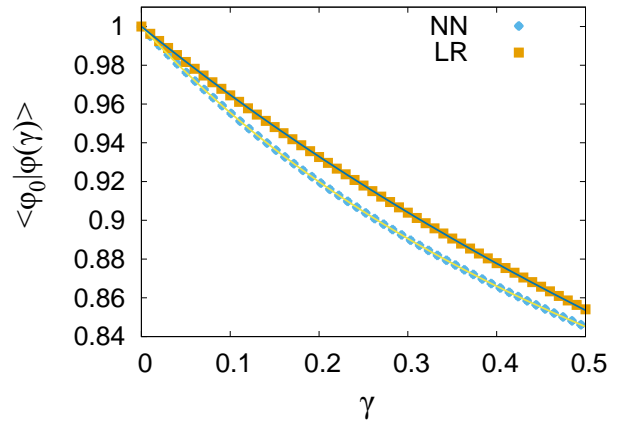}
\caption{(Colour online) Overlap of the flat-band eigenstate $\phi_{0}$ at
$\gamma=0$ with the eigenstate $\phi(\gamma)$ for $\gamma\neq 0$.
For small $\gamma$ both nearest-neighbour (NN) and long-range (LR) dipolar interaction
lead to the same functional form (\ref{eq:overlap}).}
\label{fig:overlap}
\end{figure}

We have performed the diagonalizaton of the spin-wave Hamiltonian for
nearest-neighbour dipolar interactions and for the energy of the
flat-band to obtain expression (\ref{eq:freq}).  For the long-range dipolar
interactions we diagonalize spin-wave Hamiltonian numerically.

The surprising finding is that including the long-range terms has barely any
effect on the flat band. Fig.~\ref{fig:spin_wave_spectrum_lr} shows the full
spin-wave spectrum for the case of long-range dipolar interactions with
$\gamma=0.1 J$. While the size of the  gap is a bit reduced by the inclusion of
these terms, the scaling of the gap with $\gamma$ remains the same,
$\Delta\propto \sqrt{\gamma}$ -- see Fig. \ref{fig:spin_gap_scaling}. While we
understand this scaling analytically for the nearest-neighbour model -- see
Eq.~(\ref{eq:freq}) -- its insensitivity to the long-range terms may appear
puzzling.

In order to understand the reason, it is instructive to take another look
at the microscopic picture of the weathervane mode developed in
Section~\ref{sec:local_mode} for the nearest-neighbour case. Two observations
are in order. Firstly, the same logic that we used to argue that spin
$\mathbf{S}_0$ in Fig.~\ref{fig:dipolar_gs} was unaffected if two of its
neighbours, $\mathbf{S}_1$ and $\mathbf{S}_2$, tilted away from their
respective equilibrium positions by  opposite angles also tells us that
$\mathbf{S}_0$ remains unaffected by the weathervane mode even in the presence
of longer-range terms. Each pair of equidistant spins (out of the six spins
participating in the weathervane mode) generates no torque on $\mathbf{S}_0$
since the two spins tilt in  opposite directions. As for the spins further
away from the hexagon hosting this mode, we notice that the six oscillating
spins still always add to zero. This implies that the spin dipole moment
associated with the weathervane mode is zero and hence the leading term in the
multipole expansion describing the interaction of other spins with the
weathervane mode is at best quadrupolar and hence rapidly decays with distance.
Therefore it is natural to expect that the weathervane mode remains essentially
local which, in turn, explains the observed flatness of the band. The main
effect of the longer-range rerms will be to renormalise couplings in the EOMs
such as Eq.~(\ref{eq:EOM_lin}) while preserving their structure.

We also consider the overlap of the eigenstates in the zero $\phi_{0}$ and
finite-energy $\phi(\gamma)$ flat-bands as $\gamma$ increases. In Fig.
\ref{fig:overlap} overlaps for nearest-neighbour dipolar interactions and
long-range dipolar interactions are plotted. In both cases we observe a
continuous decrease of the overlap with the increase of $\gamma$ with both
curves fitting well to a functional form
\begin{equation}
\langle \phi_{0}|\phi(\gamma) \rangle=\frac{1}{\sqrt{1+a\gamma/\left(J+b\gamma/2\right)}},
\label{eq:overlap}
\end{equation}
where in the long-range case $a=0.75$ and $b=0$ are renormalisation factors
of dipolar interactions in $x$ and $y$ components of the flat-band eigenstates.
For the nearest-neighbour dipolar interactions both parameters $a$ and $b$ are
equal to $1$ and equation above can also be obtained from the equations of
motion Eq.~(\ref{eq:EOM_lin}).

\section{Conjugate variables representation and Maxwellian counting}
In this section, we make contact with the lore on excitations and frustrated magnets, in particular the ideas of
constraint counting used to derive the size of the zero-energy space of excitations.

In the absence of dipolar
interactions, the origin of the flat band at zero energy is understood by noting that,
for $n$-component spins represented by unit length vectors,
the number of degrees of freedom per unit cell of three spins is $D_n=3(n-1)$. At the same time, the
constraints per unit cell imposed by the Hamiltonian are evaluated as $K_n=2n$, as two momentless
triangles inhabit each unit cell, and momentlessness requires one constraint on each of the $n$ spin components.

For Heisenberg spins, these are balanced, $D_3=K_3$, while for
for XY spins, $n=2$, there exist ground states satisfying one more constraint
than they use degrees of freedom: $D_2-K_2=-1$. For the subset of coplanar ground states, an unconstrained
degree of freedom therefore remains, as there are three out of plane degrees of freedom for the two remaining
constraints on the total spin of the two triangles in the unit cell.

The spin wave spectrum for three spins consists of three bands, as pairs of conjugate variables, $p_\eta(k),q_\eta(k)$
for each of the three modes $\eta=1\ldots3$ at wavevector $k$ appear in the  canonical Hamiltonian:
\begin{equation}
H=\sum_{\eta,k}\alpha_\eta(k)p^2+\beta_\eta(k)q^2\ .
\end{equation}
The spin-wave frequencies are thus computed similarly to harmonic oscillator
modes
\begin{equation}
\omega_{\eta}(k)=\sqrt{\alpha_{\eta}(k)\beta_{\eta}(k)}\ .
\end{equation}
The underconstraint identified above translates into the vanishing of one  band of coefficients (out of six), say of all
\begin{equation}
\alpha_{1}(k) \equiv 0\ .
\label{eq:zero}
\end{equation}

From this perspective, the possibility of lifting the mode to a flat finite frequency band looks outlandish -- for a vanishing
frequency, Eq.~\ref{eq:zero} imposes no constraints on the behaviour of $\beta_1(k)$ in order to satisfy
$\omega_1(k)=\sqrt{\alpha_{1}(k)\beta_{1}(k)}\equiv0$. However, for $\omega_1(k)=\sqrt{\alpha_{1}(k)\beta_{1}(k)}\equiv \Omega>0$,
one requires the momentum dependence of $\beta_1(k)=\Omega^2/\alpha_{1}(k)$ to be the inverse of that of $\alpha_{1}(k)$.

The way this is resolved is that both $\alpha_{1}(k)$ and $\beta_1(k)$ are
constant functions of $k$, related to the hexagonal motifs of the weathervane
modes: they correspond to the two (in- and out-of-plane) components of the spin
wave. The spin wave mode corresponds to an excitation of these components with
the the relative phase shifted by $\pi/2$. This describes elliptical precession
of spins with an eccentricity discussed above.

This `doubled' flat band structure of the kagome Heisenberg excitations has, to
our knowledge, not yet been identified even for the much-studied pure
nearest-neighbour Heisenberg model. It is a remarkable fact that it remains
stable to the addition of the dipolar interactions and manifests itself in the
band flatness at a finite frequency $\Omega>0$.

\begin{figure}[hbt]
\begin{center}
\raggedleft
\includegraphics[trim=1.cm 1.cm 1.cm 2.cm, clip=true,height=0.17\textwidth]{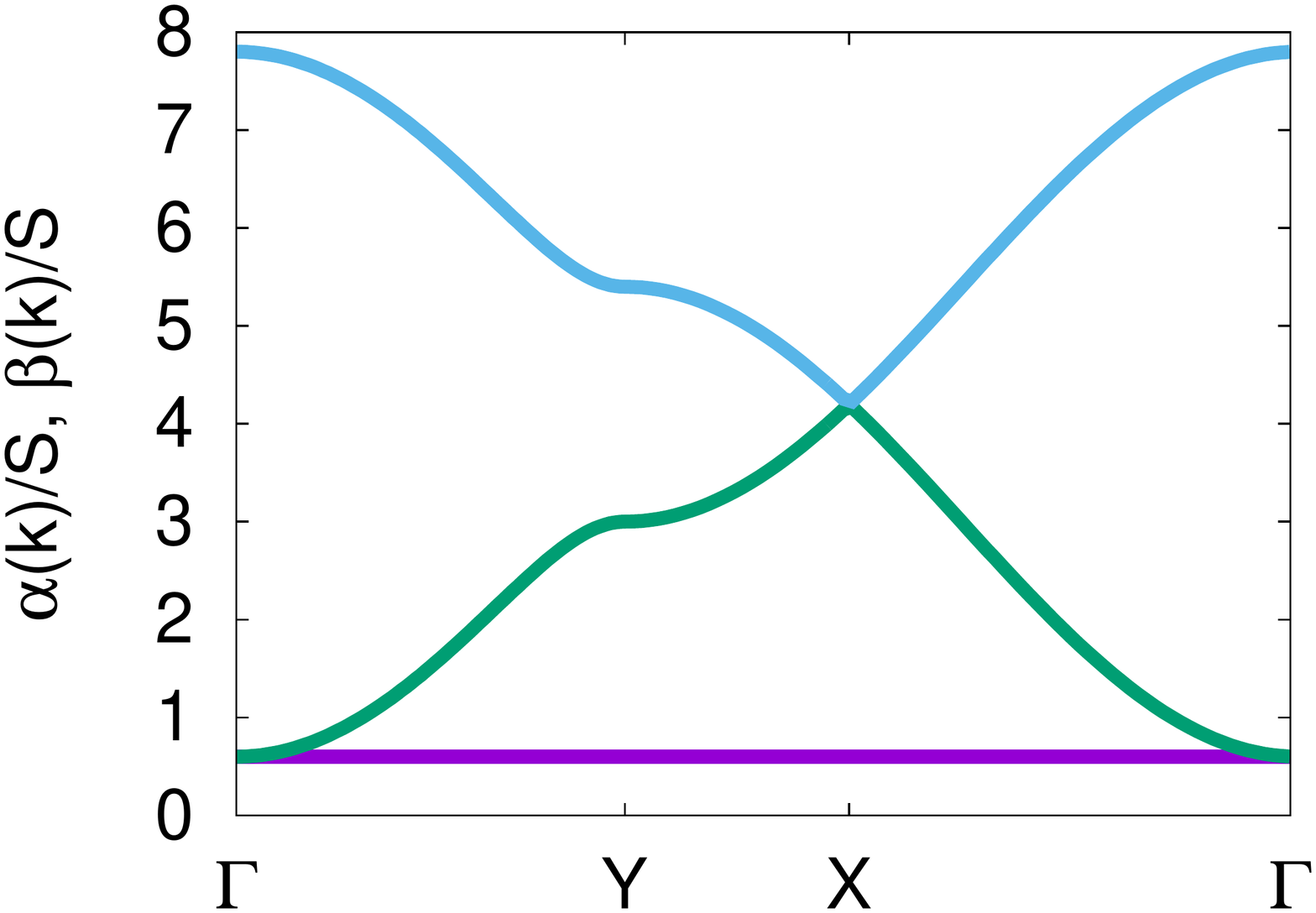}
\includegraphics[trim=6.cm 1.cm 2.cm 2.cm, clip=true,height=0.17\textwidth]{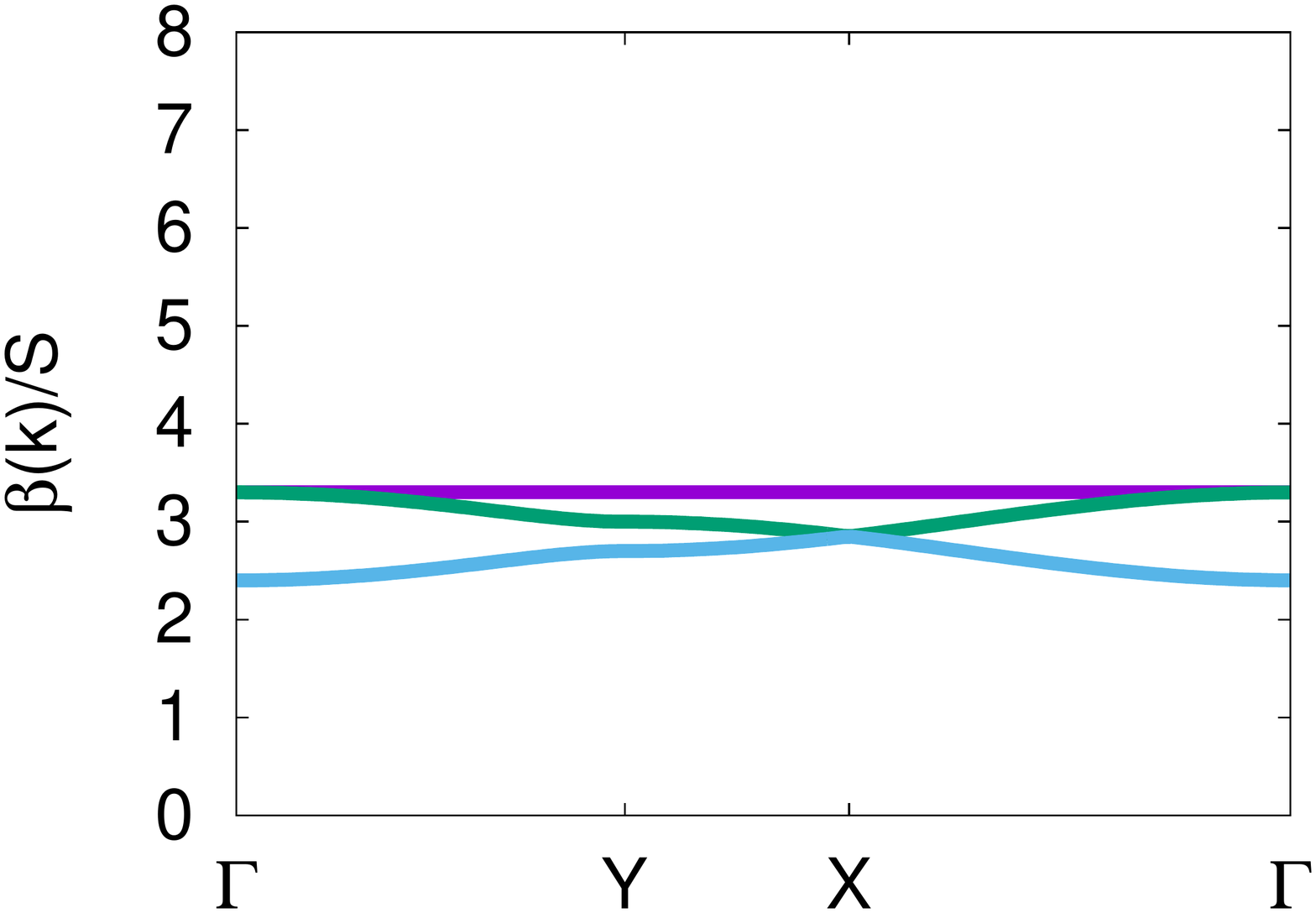}
\includegraphics[trim=2.2cm 2.cm 2.5cm 2.cm, clip=true,width=0.46\textwidth]{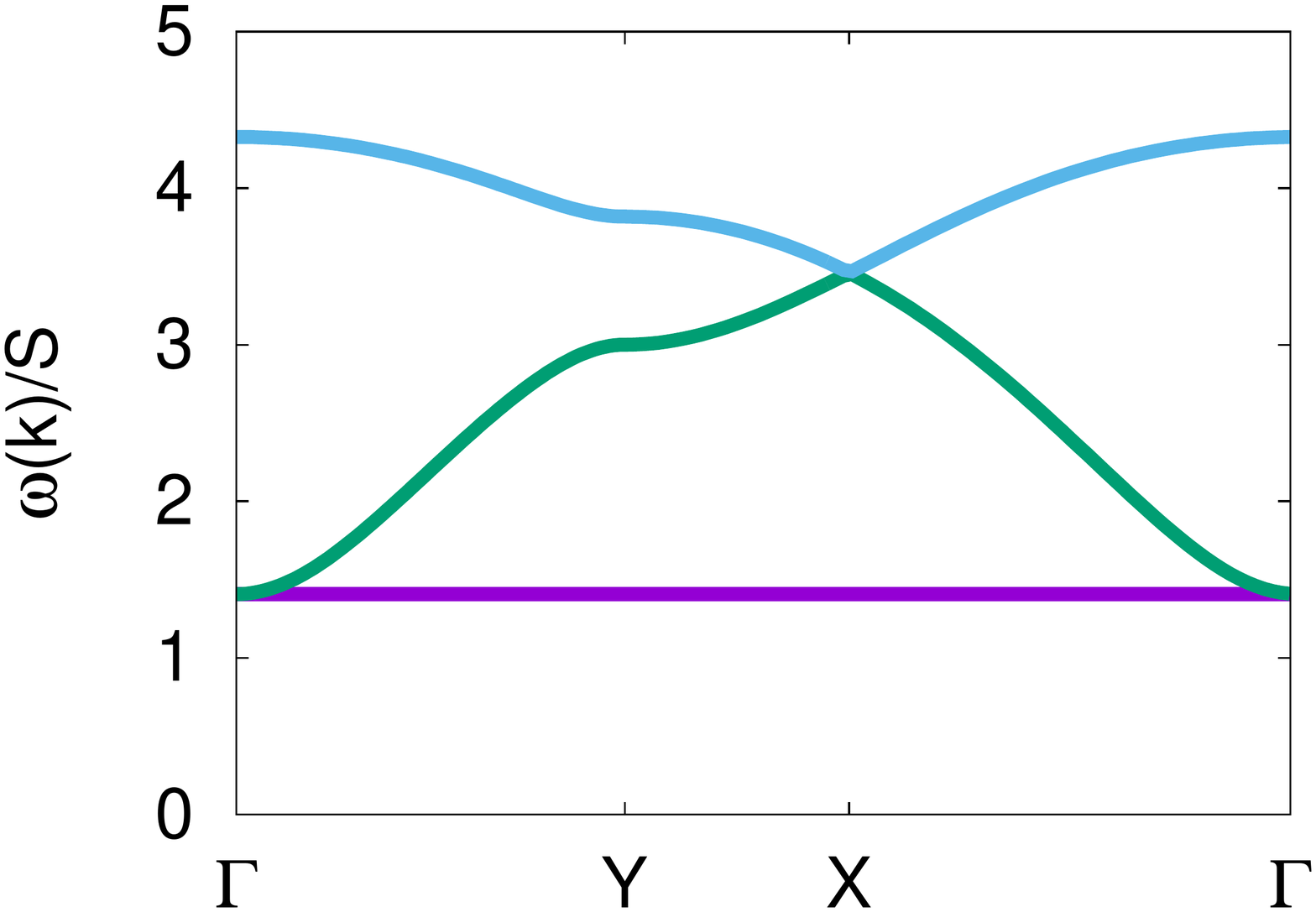}
\caption{(Colour online) Spectra of conjugate variables coefficients
$\alpha$ (left) and $\beta$ (right) for nearest-neighbour dipolar interactions ${\gamma=0.2 J}$.
Full spin-wave spectrum constructed from $\alpha$ and $\beta$ as ${\omega=\sqrt{\alpha \beta}}$.}
\label{fig:alpha_beta_spectrum}
\end{center}
\end{figure}

\section{Extensions to other systems}
\label{sec:extensions}

\subsection{Dzyaloshinskii--Moriya interactions}
Having established the existence of weathervane modes and corresponding flat
magnon bands for KHAFM with dipolar interactions, a natural question is whether
this phenomenon is confined to this specific model. In this section we show
that a number of extensions and modifications of our model preserve the
flatness of the magnon band. The first, most straightforward generalisation of
our model involves additional Dzyaloshinskii--Moriya (DM) interactions $\delta
H_\text{DM} = \sum_{\langle i,j\rangle} \mathbf{D}_{ij} \cdot
\left(\mathbf{S}_i \times \mathbf{S}_j\right)$, which should generically be
present in kagom\'{e}
systems~\cite{Dzyaloshinskii1957,Dzyaloshinsky1958,Moriya1960,
Yildirim1994,Yildirim1995,Elhajal2002,Ballou2003,Cepas2008,Chernyshev2015} due
to the lack of the inversion symmetry with respect to the mid-points of its
bonds. These interactions are known to exist in materials with 2D kagom\'{e}
planes such as the herbertsmithite~\cite{Zorko2008} and
jarosite~\cite{Matan2006,Yildirim2006} compounds. For a strictly 2D kagom\'{e}
system (or if the kagom\'{e} plane is also a mirror plane), vectors
$\mathbf{D}_{ij}$ must be strictly perpendicular to the plane. Moreover, if the
inversion symmetry with respect to the \emph{sites} of the lattice is not
broken, the DM coupling parameter must be uniform,
$\mathbf{D}_{ij}=-D\hat{\mathbf{x}}$, provided that all pairs $\langle
i,j\rangle$ are ordered in an anticlockwise manner around lattice
triangles.~\footnote{Here we adopt the site ordering convention of
Refs.~[\onlinecite{Elhajal2002}] and [\onlinecite{Ballou2003}]; other authors
prefer ordering them along the 1D chains of spins, which results in alternating
directions of $\mathbf{D}_{ij}$ between up- and down-triangles -- see e.g.
Refs.~[\onlinecite{Cepas2008}], [\onlinecite{Chernyshev2015}].}
Therefore, as long as $D>0$ (using the aforementioned sign convention), the DM
interactions are not frustrating: by themselves, they stabilise a $120^\circ$
state with positive spin vorticity, and the two ground states of
Hamiltonian~(\ref{eq:dipolar_hamiltonian}) are already of this kind. This could
be seen explicitly using the alternative expression for the Hamiltonian
consisting of both Heisenberg and nearest-neighbour dipole--dipole interactions
derived in Appendix~\ref{sec:dipolar_alt}. With the addition of the DM
interactions, the combined Hamiltonian becomes
\begin{multline}
H_\text{KH+NND+DM} = \sum_\vartriangle \left\{\frac{J+\gamma}{2}\, \mathbf{S}_\vartriangle^2
\right.\\
-\gamma\mathbf{S_\vartriangle}\cdot\sum_{i\in\vartriangle} \hat{\mathbf{e}}_{i}
\left(\mathbf{S}_i\cdot \hat{\mathbf{e}}_{i}\right)
+ \gamma \left[\sum_{i\in\vartriangle}\mathbf{S}_i\cdot \hat{\mathbf{e}}_{i}\right]^2\\
\left.
- \left(D+\frac{\sqrt{3}}{2} \gamma\right)\sum_{\langle i\hookrightarrow j\rangle\in\vartriangle}
\left(\mathbf{S}_{i}\times\mathbf{S}_j\right)\!\cdot
\hat{\mathbf{x}}\right\}.
 \label{eq:alt_triang_ham_with DM}
\end{multline}
The inclusion of the DM term simply makes the coefficient in front of the last
term bigger, and the magnitude of that term was already saturated by the ground
state shown in Fig.~\ref{fig:dipolar_gs}. Moreover, this equation shows that as
long as the antiferromagnetic Heisenberg coupling $J$ is strong enough to
stabilise $120^\circ$ ground states, the spin configuration of
Fig.~\ref{fig:dipolar_gs} remains the ground state even for moderately negative
DM coupling $-\sqrt{3}\gamma/2 \leq D <0$ which, without the dipolar term,
would stabilise the $120^\circ$ state of the opposite spin vorticity. (The only
frustrating term in Eq.~(\ref{eq:alt_triang_ham_with DM}) is the second one,
and it remains zero in all $120^\circ$ configurations).

\begin{figure}[hbt]
\begin{center}
\includegraphics[trim=2.2cm 2.cm 2.5cm 2.cm, clip=true,width=0.40\textwidth]{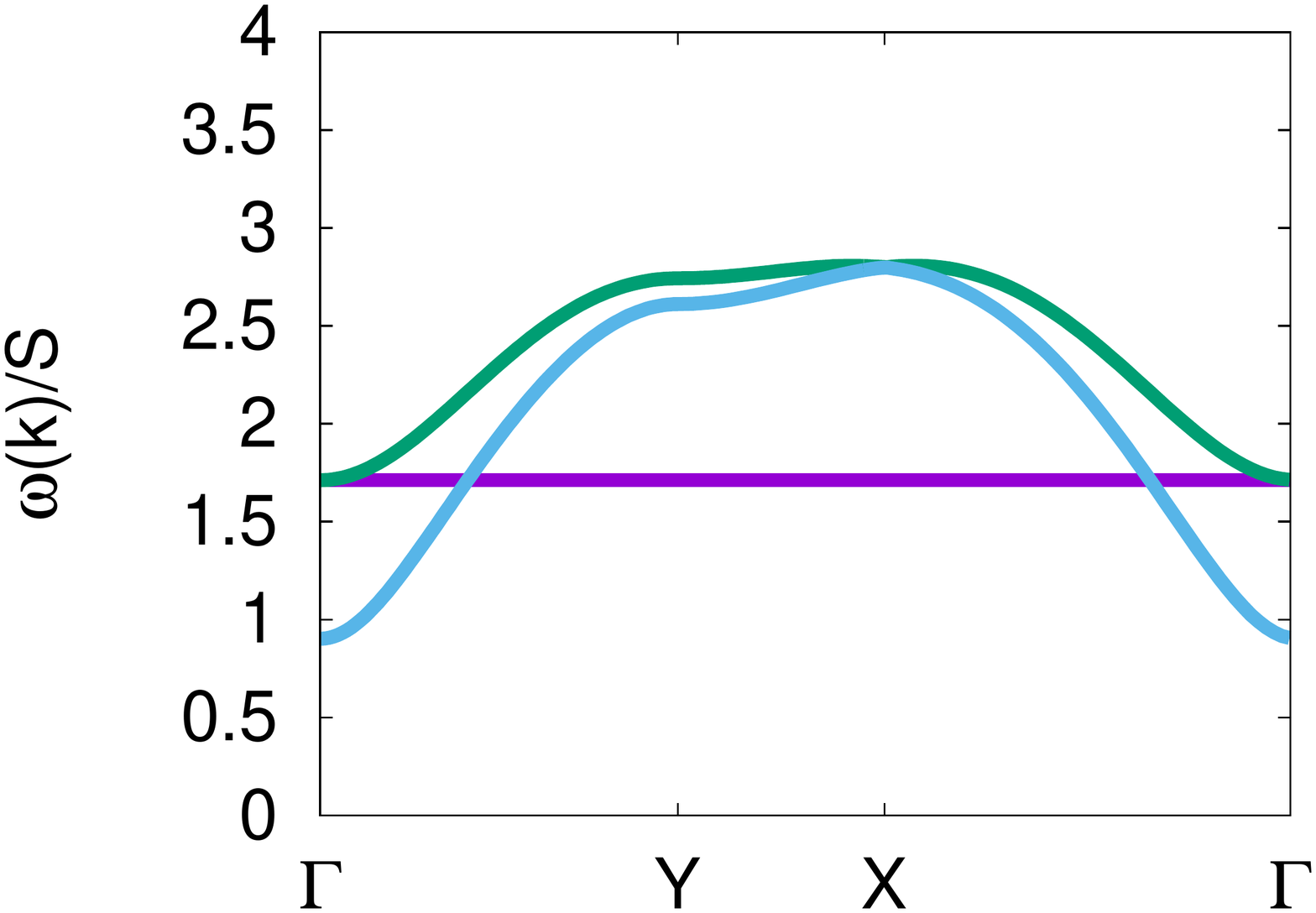}
\includegraphics[trim=2.2cm 2.cm 2.5cm 2.cm, clip=true,width=0.40\textwidth]{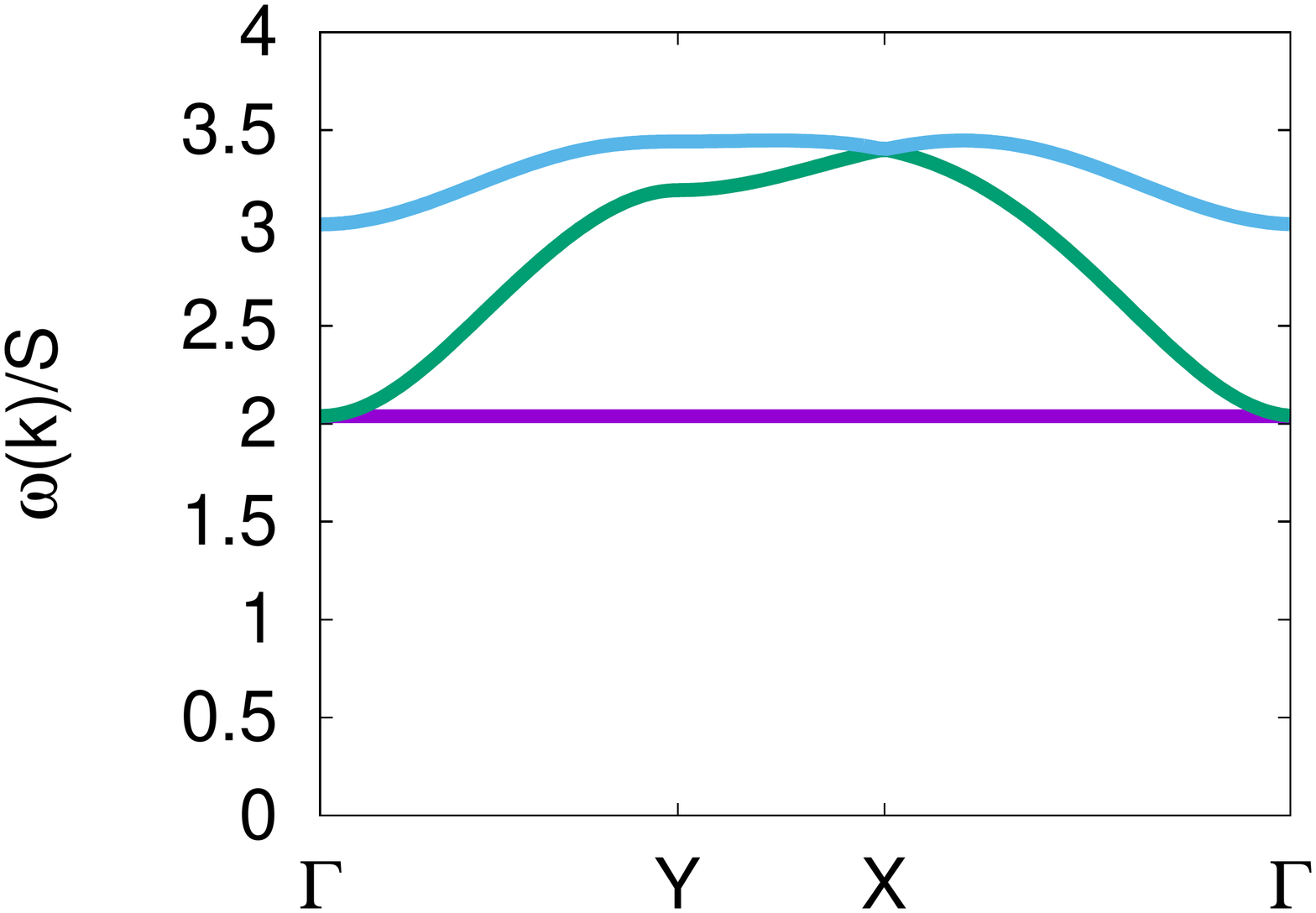}
\caption{(Colour online) Full spin-wave spectrum for $J=1$, $D=0.2$ and $\gamma=0.01$ (top),  $\gamma=0.1$ (bottom).}
\label{fig:DM_spectrum}
\end{center}
\end{figure}

A straightforward modification of the linearised EOMs~(\ref{eq:EOM_lin}) for
the same weathervane mode now reads
\begin{eqnarray}\label{eq:EOM_lin_DM}
  \frac{\hbar}{S} \frac{d S^{x}_{2}}{d t} &=&  3\left( J+\frac{\gamma}{2}+ \sqrt{3}D\right) S^{y}_{2},
  \nonumber \\
  \frac{\hbar}{S} \frac{d S^{y}_{2}}{d t} &=&  - \left(3\gamma +2\sqrt{3}D\right) S^{x}_{2}.
\end{eqnarray}
The resulting frequency is
\begin{equation}\label{eq:freq_DM}
  {\hbar}\omega =  3 S\sqrt{\left(\gamma+ \frac{2\sqrt{3}}{3}D\right)\left( J+\frac{\gamma}{2}+ \sqrt{3}D\right)},
\end{equation}
which, in the absence of dipolar interactions $\gamma=0$, reduces to the
frequency of the flat mode found in Ref.~[\onlinecite{Chernyshev2015}]. Note,
however, that while the addition of DM interactions (with all $\mathbf{D}_{ij}$
parallel to one another) reduces the symmetry of the Heisenberg Hamiltonian
from SU(2) to U(1), the Goldstone theorem still guarantees the existence of a
gapless mode, and a linearly-dispersive (at small $k$) mode was indeed found in
in Ref.~[\onlinecite{Chernyshev2015}]. The weathervane mode, while flat, is not
the lowest energy excitation in this case. This situation is changed
dramatically in the presence of dipolar interactions, which further reduce the
symmetry to $\mathbb{Z}_2$ thus completely gapping the spin waves.
While for small $\gamma \ll D$ formerly gapless dispersive mode remains below that of the flat
band for sufficiently strong dipolar term the
flat-band can become again the lowest band in the spectrum (See Figure \ref{fig:DM_spectrum}).

Note that irrespective of a particular choice of parameters, the flat band is
always touched by another band, in full agreement with the counting argument of
Ref.~[\onlinecite{Bergman2008}].

\subsection{Dipolar-like interactions with arbitrary sign}

Further generalisations of the model include considering the dipolar-type term
with the negative coupling constant $\gamma<0$. We remind the reader that the
coupling constant of the genuine dipole--dipole interaction is fixed:
$\gamma_\text{dp} = g^2 \mu_\text{B}^2/a^3$. However, a term of this type need
not arise from dipole--dipole interactions. As was shown by
Moriya~\cite{Moriya1960} and further elaborated by Yildirim et
al.~\cite{Yildirim1994,Yildirim1995}, the superexchange mechanism in the
presence of spin-orbit interactions leads to the spin--spin interaction of the
general form
\begin{equation}\label{eq:superechange}
  H_{ij}= J_{ij} \mathbf{S}_i\cdot \mathbf{S}_j + \mathbf{D}_{ij}
\cdot \left(\mathbf{S}_i \times \mathbf{S}_j\right) + \sum_{\alpha,\beta}
{S}_i^\alpha \Gamma_{ij}^{\alpha\beta}{S}_j^\beta
\end{equation}
where the first two terms describe the familiar Heisenberg and
Dzyaloshinskii--Moriya interactions while the last term is the anisotropic
exchange term characterised by a symmetric traceless tensor
$\Gamma^{\alpha\beta}$. Note that the dipolar coupling in
Eq.~(\ref{eq:int_matrix}) is exactly of this type, but whenever this term is
generated by superexchange rather than actual dipole--dipole interactions, the
sign of $\gamma$ can be arbitrary. Moreover, if we insist on the full set of
symmetries of a kagom\'{e} plane, we can argue that such dipolar-like
interaction is just one of the two couplings consistent with these symmetries.
Specifically, note that a real symmetric traceless tensor
$\Gamma^{\alpha\beta}$ has five independent components. For the purpose of this
argument let us choose an orthogonal set of axes for a particular bond as
follows: let the $z$- axis align with the bond, the $y$-axis lie in the
kagome\'{e} plane and the $x$-axis be perpendicular to it (in accordance with
the choice of the $x$-axis direction made earlier\footnote{The choice of the
$x$-axis for the out-of-plane direction is a consequence of aligning local
$z$-axes along each spin in their equilibrium state -- a standard choice for
the Holstein--Primakoff transformations in the spin-wave theory. This makes
referring to the XXZ anisotropy later in text somewhat awkward, but we prefer
it to changing the axis labelling convention halfway through the manuscript.}).
The symmetries of the kagom\'{e} plane include $(y,z)$ and $(x,y)$ mirror
planes, with the former coinciding with the kagome\'{e} plane itself and the
latter being perpendicular to the bond at its midpoint. As a result, the
Hamiltonian should be invariant under $x\to -x$ and $z\to -z$ transformations
(with the components of spins transforming accordingly). Therefore
$\Gamma^{xy}=\Gamma^{xz}=\Gamma^{yz}=0$, leaving us with just two independent
parameters. Choosing $\gamma$, the dipolar-like coupling, to be one of them,
the other parameter is $3\delta = \Gamma^{yy} - \Gamma^{xx}$, the difference
between in-plane and out-of-plane couplings for spin components orthogonal to
the bond.

Therefore, the most general bilinear spin-spin interaction consistent with the
symmetries of the kagom\'{e} lattice can be written as
\begin{multline}
\label{eq:general_hamiltonian}
  H_{ij}= (J+\gamma+\delta) \mathbf{S}_i\cdot \mathbf{S}_j + \mathbf{D}_{ij}
\cdot \left(\mathbf{S}_i \times \mathbf{S}_j\right)\\
-3 \gamma \left(\mathbf{S}_i\cdot \hat{\mathbf{r}}_{ij}\right)
\left(\mathbf{S}_j\cdot \hat{\mathbf{r}}_{ij}\right) -3 \delta S_i^x S_j^x,
\end{multline}
where $\mathbf{D}_{ij}$ is a vector in the $\mathbf{\hat{x}}$ (i.e., out-of-plane) direction
with the aforementioned sign convention. Given the superexchange origin of
these terms, the signs of coupling constants $\gamma$ and $\delta$ can now be
arbitrary, and we should only concern ourselves with the nearest-neighbour
interactions since this mechanism is exponentially suppressed with distance.

Let us first consider the case of $\gamma<0$ and $\delta=0$. While it may not
immediately obvious what the ground sate(s) of
Hamiltonian~(\ref{eq:nn_dipolar_hamiltonian}) or, equivalently
(\ref{eq:alt_triang_ham}) with $\gamma<0$ might be -- after all, the individual
terms in Eq.~(\ref{eq:alt_triang_ham}) are now minimised by different spin
configurations -- it turns out that the energy in minimised by the
all-in/all-out $120^\circ$ states where all $\mathbf{S}_i
\parallel \hat{\mathbf{e}}_{i}$. In other words, the two ground states are
obtained from the ground states in the positive $\gamma$ case by rotating all
spins by $\pi/2$. In fact, the Hamiltonian with the negative ``dipolar''
coupling $\gamma<0$ is \emph{less frustrated} than the identical Hamiltonian
with $\gamma>0$ due to the dominant nature of the third term in
Eq.~(\ref{eq:alt_triang_ham}); the ground state energy for $\gamma<0$ is $1.5
\lvert \gamma\rvert$ lower than that for $\gamma>0$ of the same magnitude. As a
result, even a weak \emph{ferromagnetic} Heisenberg coupling $J<0$ does not
immediately destabilise this state: As long as $\lvert J\rvert < \lvert \gamma
\rvert/2$, the ground state remains the all-in/all-out $120^\circ$ state. For
strong ferromagnetic coupling $\lvert J\rvert
> \lvert \gamma \rvert/2$ the energy is minimised when all three spins are parallel
to one another and perpendicular to the kagom\'{e} plane.

An additional DM term with $D>0$ can only further stabilise the all-in/all-out
$120^\circ$ state as it would counteract the frustrating effect of the last
term in Eq.~(\ref{eq:alt_triang_ham}), which by itself would favour the state
with the opposite spin vorticity.

Without the DM term, the linearised EOMs for the weathervane mode become
\begin{equation}\label{eq:EOM_negative_D}
  \frac{\hbar}{S} \frac{d S^{x}_{2}}{d t} =  3\left( J+\frac{3\lvert \gamma \rvert}{2}\right) S^{y}_{2},
  \quad
  \frac{\hbar}{S} \frac{d S^{y}_{2}}{d t} =  - 9\lvert \gamma \rvert S^{x}_{2},
\end{equation}
and the frequency of such a mode is now given by
\begin{equation}\label{eq:freq_negative_D}
  {\hbar}\omega =  9 S\sqrt{ \lvert \gamma \rvert\left( \frac{J}{3}+\frac{\lvert \gamma \rvert}{2}\right)}.
\end{equation}
For small $\lvert \gamma \rvert\ll J$ it is still proportional to the square
root of the coupling constant $\lvert \gamma \rvert$, but it is higher than
that given by Eq~(\ref{eq:freq}) for $\gamma>0$ of the same magnitude.
 This is not surprising, since we already saw that the system is less
frustrated for $\gamma<0$ and has deeper energy minima.

Modifications of Eqs.~(\ref{eq:EOM_negative_D}) and (\ref{eq:freq_negative_D})
for the case of additional DM interactions are straightforward since the
contributions of the DM terms into the effective magnetic field acting on a
given spin are invariant under the rotations of all spins by the same angle in
the plane of the lattice and hence they contribute to the EOMs in exactly the
same way they do for the $\gamma>0$ case (see
Eqs.~(\ref{eq:EOM_lin_DM},\ref{eq:freq_DM})):
\begin{eqnarray}\label{eq:EOM_neg_D_DM}
  \frac{\hbar}{S} \frac{d S^{x}_{2}}{d t} &=&  3\left( J+\frac{3\lvert \gamma\rvert}{2}+ \sqrt{3}D\right) S^{y}_{2},
  \nonumber \\
  \frac{\hbar}{S} \frac{d S^{y}_{2}}{d t} &=&  - \left(9\lvert \gamma\rvert +2\sqrt{3}D\right) S^{x}_{2}.
\end{eqnarray}
The frequency of such mode becomes
\begin{equation}\label{eq:freq_neg_D_DM}
  {\hbar}\omega =  3 S\sqrt{\left(3\lvert \gamma\rvert + \frac{2\sqrt{3}}{3}D\right)
  \left( J+\frac{3\lvert \gamma\rvert}{2}+ \sqrt{3}D\right)}.
\end{equation}

\subsection{XXZ anisotropy}

Finally, we turn our attention to the case of $\delta\neq 0$ (see
Eq.~(\ref{eq:general_hamiltonian})). Working out the full phase diagram of this
model is beyond the scope of our paper. While strong out-of-plane coupling can
easily cant the spins, it has been found that in the absence of dipolar-like
terms ($\gamma=0$), the ground states of (\ref{eq:general_hamiltonian}) for
reasonably small positive values of $\delta \in (0,J/2)$ and arbitrarily small
DM interactions with $D>0$ are $120^\circ$ states with positive spin
vorticity~\cite{Essafi2016}. Moreover, such ground states are stabilised for
any value of $\delta$ (positive or negative) by a sufficiently strong DM term.
As we have argued, the presence of short-range dipolar-like terms ($\gamma\neq
0$) would then merely break the remaining global O(2) symmetry down to
$\mathbb{Z}_2$. The upshot is that as long as the ground states of our system
are either the ones described in Section~\ref{sec:dipolar_interactions}, or the
all-in/all-out states described in this section, the above analysis applies.

Therefore, for $\gamma>0$, the linearised EOMs for the weathervane mode become
\begin{eqnarray}\label{eq:EOM_lin_DM_xxz}
  \frac{\hbar}{S} \frac{d S^{x}_{2}}{d t} &=&  3\left( J+\frac{\gamma}{2}+ \delta +\sqrt{3}D\right) S^{y}_{2},
  \nonumber \\
  \frac{\hbar}{S} \frac{d S^{y}_{2}}{d t} &=&  - \left(3\gamma +6\delta +2\sqrt{3}D\right) S^{x}_{2}.
\end{eqnarray}
The resulting frequency is
\begin{equation}\label{eq:freq_DM_xxz}
  {\hbar}\omega =  3 S\sqrt{\left(\gamma+ 2\delta+ \frac{2\sqrt{3}}{3}D\right)\left( J+\frac{\gamma}{2}+ \delta+ \sqrt{3}D\right)}.
\end{equation}

Meantime, for $\gamma<0$, Eqs.~(\ref{eq:EOM_neg_D_DM}) and
(\ref{eq:freq_neg_D_DM}) become generalised to
\begin{eqnarray}\label{eq:EOM_neg_D_DM_xxz}
  \frac{\hbar}{S} \frac{d S^{x}_{2}}{d t} &=&  3\left( J+\frac{3\lvert \gamma\rvert}{2}+ \delta+ \sqrt{3}D\right) S^{y}_{2},
  \nonumber \\
  \frac{\hbar}{S} \frac{d S^{y}_{2}}{d t} &=&  - \left(9\lvert \gamma\rvert+ 6\delta +2\sqrt{3}D\right) S^{x}_{2}.
\end{eqnarray}
The frequency of the weathervane mode becomes
\begin{equation}\label{eq:freq_neg_D_DM_xxz}
  {\hbar}\omega =   3 S\sqrt{\left(3\lvert \gamma\rvert + 2\delta + \frac{2\sqrt{3}}{3}D\right)
  \left( J+\frac{3\lvert \gamma\rvert}{2}+\delta+ \sqrt{3}D\right)}.
\end{equation}

We conclude that the flat magnon band is a very generic feature of the
classical kagom\'{e} antiferromagnets. It is very robust and its nature is
largely independent of the details of the interactions, as long as their net
effect is to stabilise one of the $120^\circ$ planar ground states with
positive spin vorticity.

\subsection{The fate of the Goldstone mode}

It is interesting to note that in the absence of dipolar-like terms in the
Hamiltonian given by Eq.~(\ref{eq:general_hamiltonian}), either
Dzyaloshinskii--Moriya ($D\neq 0$) or XXZ anisotropy ($\delta \neq 0$) are
sufficient to lift the flat band to a finite energy\cite{Chernyshev2015}, as
can be clearly seen from Eqs.~(\ref{eq:freq_DM_xxz}) and
(\ref{eq:freq_neg_D_DM_xxz}). Nevertheless, both of these terms keep the full
spin wave spectrum gapless: they only break the SU(2) symmetry of the
Heisenberg Hamiltonian down to U(1), and the Goldstone theorem still guarantees
the existence of a gapless mode with $\omega_{q\to 0}\to 0$. A dipolar-like
term in Eq.~(\ref{eq:general_hamiltonian}) is the only term that breaks the
symmetry of the Hamiltonian down to $\mathbb{Z}_2$ and thus  opens the gap in
the spin wave spectrum. A straightforward analysis of uniform deviations of all
spins from their equilibrium positions for the case of $\gamma>0$ yields the
following equations of motion:
\begin{eqnarray}\label{eq:EOM_Goldstone}
  \frac{\hbar}{S} \frac{d S^{x}}{d t} &=&  12\gamma S^{y},
  \nonumber \\
  \frac{\hbar}{S} \frac{d S^{y}}{d t} &=&  - \left(6 J + 9 \gamma + 6\delta +2\sqrt{3}D\right) S^{x}.
\end{eqnarray}
The frequency of this mode becomes
\begin{equation}\label{eq:freq_Goldstone}
  {\hbar}\omega =   6 S\sqrt{\gamma\left(2J + 3 \gamma + 2\delta + \frac{2\sqrt{3}}{3}D\right)}.
\end{equation}

A similar calculation for $\gamma<0$ yields
\begin{eqnarray}\label{eq:EOM_Goldstone_wrong_sign}
  \frac{\hbar}{S} \frac{d S^{x}}{d t} &=&  12\left|\gamma\right| S^{y},
  \nonumber \\
  \frac{\hbar}{S} \frac{d S^{y}}{d t} &=&  - \left(6 J + 3 \left|\gamma\right| + 6\delta +2\sqrt{3}D\right) S^{x},
\end{eqnarray}
and consequently
\begin{equation}\label{eq:freq_Goldstone_wrong_sign}
  {\hbar}\omega =   6 S\sqrt{\left|\gamma\right|\left(2J + \left|\gamma\right| + 2\delta + \frac{2\sqrt{3}}{3}D\right)}.
\end{equation}

Therefore, for small dipolar interactions the gap in the spin wave spectrum is
always proportional to $\sqrt{\left|\gamma\right|}$, but it need not correspond
to the flat band, which may be shifted to higher energy by DM interactions or
XXZ anisotropy.

Curiously, in the absence of DM interactions or XXZ anisotropy, the uniform
mode softens at $J=-|\gamma|/2$ on the ferromagnetic side ($J<0$), i.e.
precisely at the point where the nature of the ground state changes from the
$120^\circ$ arrangement of spins ($|J|<|\gamma|/2$) to the fully-polarised
out-of-plane state ($|J|>|\gamma|/2$). This may appear puzzling since the
transition between the two ground states as a function of $J/K$ is a typical
first-order, level-crossing transition not requiring any mode softening.
However, it is easy to check that exactly at $J=\gamma/2<0$ the energy of a
uniformly canted $120^\circ$ arrangement of spins becomes independent on the
canting angle (which is consistent with the notion of a transition from the
uncanted to the maximally-canted, i.e. ferromagnetic state). It is this
degeneracy of the ground state with respect the canting angle that is reflected
in the vanishing frequency of the uniform mode.

\section{Discussion and outlook}

In summary, we have discovered a remarkable stability of the dispersionlessness
of the band of weathervane modes of the classical kagome Heisenberg
antiferromagnet. In particular, we have identified the dipolar interactions as
a particularly impressive case in point, given that it removes the continuous
Heisenberg symmetry with its concomitant gaplessness of the mode spectrum,
moving the flat band upwards along with the rest of the spectrum, while
generating only a  weak dispersion despite its long-range nature. The latter
feature we were able to connect to a Heisenberg version of the self-screening
effect found in the frustrated pyrochlore Ising system  known as dipolar spin
ice\cite{Castelnovo2012}.

More broadly, the mechanism we have identified for the persistence of the dispersionlessness
applies in a broad range of settings, including the (previously observed) cases of XXZ and
DM anisotropies, the latter of which we have discussed in a more general setting here. In the
two former cases, a combination of experimental information of the size of the gap, and the location
of the flat band (at energies above the gap) may be used to glean information about the
relative size of perturbations to the ideal Heisenberg hamiltonian.

Overall, we have found that the flat band of weathervane modes is remarkably
robust in classical kagome magnets. There are a number of further settings in
which one can study their properties, and particular scope for their
manipulation. One natural item here is the role of an applied magnetic field,
given its application connects to the well-known situation that at saturation,
flat band physics enters perhaps in its most natural way as the hopping problem
of flipped spin excitations on top of the `ferromagnetic' background, leading
to features such as a discontinuous jump in the
magnetisation~\cite{Schulenburg2002,Zhitomirsky2005,Schmidt2006,Derzhko2007a}.

While a number of distinct setups
yield dispersionless bands, the consideration of lattices such as the kagome case discussed here in detail
exhibiting
strong geometrical frustration has
long been a natural route to induce such physics.

In this respect rare-earth geometrically frustrated magnets are among the most
obvious candidates to address the flat-band effects in the spin-waves
excitation spectra. The existence of the dispersionless modes at finite energy
sharply manifest itself in inelastic neutron scattering as a finite energy,
almost ${\mathbf{k}}$-independent resonance
\cite{sw_maestro_gingras,quilliam2007_sw_therm_pyrochlore,lee_broholm_kim2000}.
A recent experimental study of the stalwart frustrated gadolinium gallium
garnet $\rm Gd_{3} Ga_{5} O_{12}$ was aimed to precisely address experimental
manifestation of a dispersionless band in inelastic neutron scattering
\cite{dAmbrumenil2015}. In this compound, magnetic ions are arranged in a 3D
hyperkagom\'{e} structure and flat-band emerges as a lowest spin-wave band
above the saturated ferromagnetic ground state in the strong magnetic field. As
this field varies it effectively plays a role of a chemical potential
controlling the population of excitations in a band. Importantly, there the
presence of dipolar interaction does not preserve the dispersionlessness to the
same degree as in the 2D kagome case, on account of the interplay of the
noncoplanarity of the triangles on the hyperkagome lattice and the
'spin--orbit' coupling of the dipolar interaction.

Regarding potential  experimental realisation of the 2D physics, recent studies
have identified a rare-earth kagom\'{e} compound $\text{Mg}_{2} \text{Gd}_{3}
\text{Sb}_{3} \text{O}_{14}$~\cite{Dun2016}. The $120^{\circ}$ ground state is
found to be stabilised by weak dipolar interactions and according to our
studies the spin-wave spectrum of the model Hamiltonian used for $\text{Mg}_{2}
\text{Gd}_{3} \text{Sb}_{3} \text{O}_{14}$ in Ref.~[\onlinecite{Dun2016}]
should contain a gapped flat-band mode, subject of course to possible
interference of other, yet to be identified, weak terms in the Hamiltonian.
Hence the presence of the flat band could be identified already at zero
magnetic field in the inelastic neutron scattering experiments and in a variety
of low-temperature behaviour of thermodynamic quantities.

On top of this, there exists  a rising number of systems which realise
kagom\'{e} lattice structure and where  dipolar interactions play important or
dominant role, such as dipolar nano-arrays \cite{Wang:2006aa,
moller2006artificial}, thin films of frustrated materials
\cite{leusink2014thin,bhuiyan2005growth,
fcc_kagome_films,irmn_long_range_order} or dipoles in optical lattices
\cite{Pupillo2009,Yao2012a,Bettles2015}. Our analysis demonstrates effects of
self-screening of dipolar interactions in such system and interesting mechanism
of lifting the formerly zero-energy flat band by squeezing the corresponding
localised modes. The investigation of strong many-body effects in such (nearly)
flat bands and their manifestation in accessible experimental probes is an
interesting direction for future research.

More broadly, dispersionless bands in the quasiparticle spectrum provide a
unique setup in which kinetic energy of corresponding modes is entirely
quenched and all the dynamics is due to disorder, interaction or quantum
statistics effects. Recent interest in flat-bands has addressed many-body
instabilities, thermodynamic effects, exotic topological phases and novel
states that could be realised there
\cite{Parameswaran2013a,Bergholtz2013,Derzhko2015}. Our study suggests flat
bands may be more stable than one might have feared.

\section{Acknowledgements} We would like thank Benoit Dou\c{c}ot, Chris Henley
and Oleg Petrenko for enlightening discussions and V.~Ravi Chandra for
collaboration on a related project. The authors would like to acknowledge the
DFG SFB 1143 grant, which provided partial support for the collaborative effort
at MPIPKS, Dresden. MM was supported in part by the ERC UQUAM grant; KS was
supported in part by the NSF DMR-1411359 grant.

\appendix
\section{Dipolar Hamiltonian on one lattice triangle}
\label{sec:dipolar_alt}

In this appendix we derive an alternative expression for the Hamiltonian
consisting of both Heisenberg and nearest-neighbour dipole--dipole interactions
on one lattice triangle. For concreteness, let us focus on the shaded triangle
hosting spins $\mathbf{S}_0$, $\mathbf{S}_1$ and $\mathbf{S}_2$ in
Fig.~\ref{fig:dipolar_gs}. We begin by introducing unit vectors
$\hat{\mathbf{e}}_{i}$ pointing from the centre of a given triangle to each of
its corners. It is easy to see that $\hat{\mathbf{r}}_{ij} =
\left(\hat{\mathbf{e}}_{j}-\hat{\mathbf{e}}_{i}\right)/\sqrt{3}$.
Straightforward but somewhat tedious algebra then yields
\begin{multline}
\frac{1}{2}\sum_{i,j\in\vartriangle}  \left(\mathbf{S}_i\cdot \hat{\mathbf{r}}_{ij}\right)
\left(\mathbf{S}_j\cdot \hat{\mathbf{r}}_{ij}\right)\\
= - \frac{1}{3}\left[\sum_{i\in\vartriangle}\mathbf{S}_i\cdot \hat{\mathbf{e}}_{i}\right]^2
+ \frac{1}{3}\mathbf{S_\vartriangle}\cdot\sum_{i\in\vartriangle} \hat{\mathbf{e}}_{i}
\left(\mathbf{S}_i\cdot \hat{\mathbf{e}}_{i}\right)\\
+ \frac{1}{6}\sum_{i,j\in\vartriangle}
\left(\mathbf{S}_i\times\mathbf{S}_{j}\right)\!\cdot\!
\left(\hat{\mathbf{e}}_{i}\times \hat{\mathbf{e}}_{j}\right)\\
= - \frac{1}{3}\left[\sum_{i\in\vartriangle}\mathbf{S}_i\cdot \hat{\mathbf{e}}_{i}\right]^2
+ \frac{1}{3}\mathbf{S_\vartriangle}\cdot\sum_{i\in\vartriangle} \hat{\mathbf{e}}_{i}
\left(\mathbf{S}_i\cdot \hat{\mathbf{e}}_{i}\right)\\
+ \frac{1}{2\sqrt{3}}\sum_{\langle i\hookrightarrow j\rangle\in\vartriangle}
\left(\mathbf{S}_{i}\times\mathbf{S}_j\right)\!\cdot
\hat{\mathbf{x}},
 \label{eq:alt_dipolar_term}
\end{multline}
where notation $\langle i\hookrightarrow j\rangle$ in the last sum indicates
anticlockwise ordering of spins around the triangle in each pair;
$\hat{\mathbf{x}}$ is the unit vector directed out of plane, consistent with
the our choice of local coordinate frames throughout this paper. The
coefficient of $1/2$ on the left-hand side of this equation is to prevent
double counting of pairs of spins.

Therefore the Hamiltonian (\ref{eq:nn_dipolar_hamiltonian}) for one lattice
triangle can be written, up to a constant,  as
\begin{multline}
H_\vartriangle = \frac{J+\gamma}{2}\, \mathbf{S}_\vartriangle^2
-\gamma\mathbf{S_\vartriangle}\cdot\sum_{i\in\vartriangle} \hat{\mathbf{e}}_{i}
\left(\mathbf{S}_i\cdot \hat{\mathbf{e}}_{i}\right)\\
+ \gamma \left[\sum_{i\in\vartriangle}\mathbf{S}_i\cdot \hat{\mathbf{e}}_{i}\right]^2
- \frac{\gamma\sqrt{3}}{2}\sum_{\langle i\hookrightarrow j\rangle\in\vartriangle}
\left(\mathbf{S}_{i}\times\mathbf{S}_j\right)\!\cdot
\hat{\mathbf{x}}.
 \label{eq:alt_triang_ham}
\end{multline}
The reason for writing the Hamiltonian in such a form is that it allows for an
easy incorporation of additional Dzyaloshiskii--Moriya terms, since the last
term of Eq.~(\ref{eq:alt_triang_ham}) is of exactly that form. The one term
whose role is not transparent is the second one. However, as long as we are
dealing with the states with $\mathbf{S}_\vartriangle=0$ (and we know this to
be true e.g., for the ground state(s) of this Hamiltonian for $J,\gamma\geq
0$), the second term vanishes. The rest of the terms are not frustrated in the
sense that the ground state minimises each of them individually for
$J,\gamma\geq 0$. This statement may not be obvious in reference to the last
term, so it might be useful to rewrite it using
\begin{equation}
\frac{\sqrt{3}}{2}\sum_{\langle i\hookrightarrow j\rangle\in\vartriangle}
\left(\mathbf{S}_{i}\times\mathbf{S}_j\right)\!\cdot
\hat{\mathbf{x}} = \frac{9}{4}\,S^2\, \mathbf{v}_\vartriangle\cdot
\hat{\mathbf{x}},
 \label{eq:vorticity term}
\end{equation}
where $\mathbf{v}_\vartriangle = {2}/{(3\sqrt{3}S^2)}\,\sum_{\langle
i\hookrightarrow j\rangle\in\vartriangle} \mathbf{S}_{i}\times\mathbf{S}_j$ is
the vector vorticity of a three-spin configuration normalised so that $\lvert
\mathbf{v}\rvert_\text{max}=1$. The vorticity is maximised by the coplanar
$120^\circ$ arrangements of spins, and $v_\vartriangle^x=1$ for the ground
states considered here.

\bibliography{dipolarref,kirref}

\end{document}